\documentclass{article}

\usepackage{times}
\usepackage{epsfig}
\usepackage{graphicx}
\usepackage{amsmath}
\usepackage{amssymb}
\usepackage{mathtools}
\usepackage{subcaption}
\usepackage{multirow}
\usepackage{booktabs}
\usepackage{breqn}
\usepackage{adjustbox}
\usepackage{pifont}
\newcommand{\cmark}{\ding{51}}%
\newcommand{\xmark}{\ding{55}}%

\begin{document}

\title{No Surprises: Training Robust Lung Nodule Detection for Low-Dose CT Scans by Augmenting with Adversarial Attacks}

\author{Siqi Liu,
Arnaud Arindra Adiyoso Setio,
Florin C. Ghesu,
Eli Gibson, \\
Sasa Grbic,
Bogdan Georgescu,
Dorin Comaniciu
\thanks{
Siqi Liu, Florin C. Ghesu, Eli Gibson, Sasa Grbic, Bogdan Georgescu and Dorin Comaniciu are with Digital Technology \& Innovation, Siemens Healthineers, Princeton, NJ, USA. (siqi.liu@siemens-healthineers.com)
\newline\newline
Arnaud Arindra Adiyoso Setio is with Digital Technology \& Innovation, Siemens Healthineers, Erlangen, Germany.
\newline\newline
The authors thank the National Cancer Institute for access to NCI’s data collected by the National Lung Screening Trial (NLST). The statements contained herein are solely those of the authors and do not represent or imply concurrence or endorsement by NCI.
\newline\newline
This manuscript has been published on IEEE Trans. on Medical Imaging.
}}


\maketitle

\begin{abstract}
Detecting malignant pulmonary nodules at an early stage can allow medical interventions which may increase the survival rate of lung cancer patients. Using computer vision techniques to detect nodules can improve the sensitivity and the speed of interpreting chest CT for lung cancer screening.
Many studies have used CNNs to detect nodule candidates.
Though such approaches have been shown to outperform the conventional image processing based methods regarding the detection accuracy, CNNs are also known to be limited to generalize on under-represented samples in the training set and prone to imperceptible noise perturbations.
Such limitations can not be easily addressed by scaling up the dataset or the models.
In this work, we propose to add adversarial synthetic nodules and adversarial attack samples to the training data to improve the generalization and the robustness of the lung nodule detection systems. 
To generate hard examples of nodules from a differentiable nodule synthesizer, we use projected gradient descent (PGD) to search the latent code within a bounded neighbourhood that would generate nodules to decrease the detector response.
To make the network more robust to unanticipated noise perturbations, we use PGD to search for noise patterns that can trigger the network to give over-confident mistakes. 
By evaluating on two different benchmark datasets containing consensus annotations from three radiologists, we show that the proposed techniques can improve the detection performance on real CT data. 
To understand the limitations of both the conventional networks and the proposed augmented networks, we also perform stress-tests on the false positive reduction networks by feeding different types of artificially produced patches.
We show that the augmented networks are more robust to both under-represented nodules as well as resistant to noise perturbations.
\end{abstract}



\section{Introduction}
Lung cancer is the leading cause of all cancer deaths \cite{siegel2019cancer}.
Detecting malignant pulmonary nodules at an early stage can allow medical interventions which may increase the survival rate of lung cancer patients.
Early-stage cancer generally manifests in the form of pulmonary nodules which are defined as rounded opacity, well or poorly defined, measuring up to 30mm in diameter \cite{hansell2008fleischner}. 
Based on the findings of the National Lung Screening Trial (NLST), the U.S. Centers for Medicare and Medicaid Services (CMS) approved screening for lung cancer of high-risk subjects to be fully reimbursed by insurance companies.
In NLST the low-dose screening test involved an approximate dose of 2 mSv, whereas full-chest CT scanning that was the major diagnostic study used to follow up nodules, involved a dose of about 8 mSv \cite{mccunney2014radiation}.
The NELSON trial also reported reduced 10 year lung-cancer mortality with CT screening with a randomized trial involving 15789 patients \cite{nelson}.
However, given the sizeable eligible screening population (8.6 million in the US) and the time cost of interpreting 3D chest CT, it substantially increases the efforts for radiologists.

\begin{figure}[t]
    \centering
    \includegraphics[width=1\linewidth]{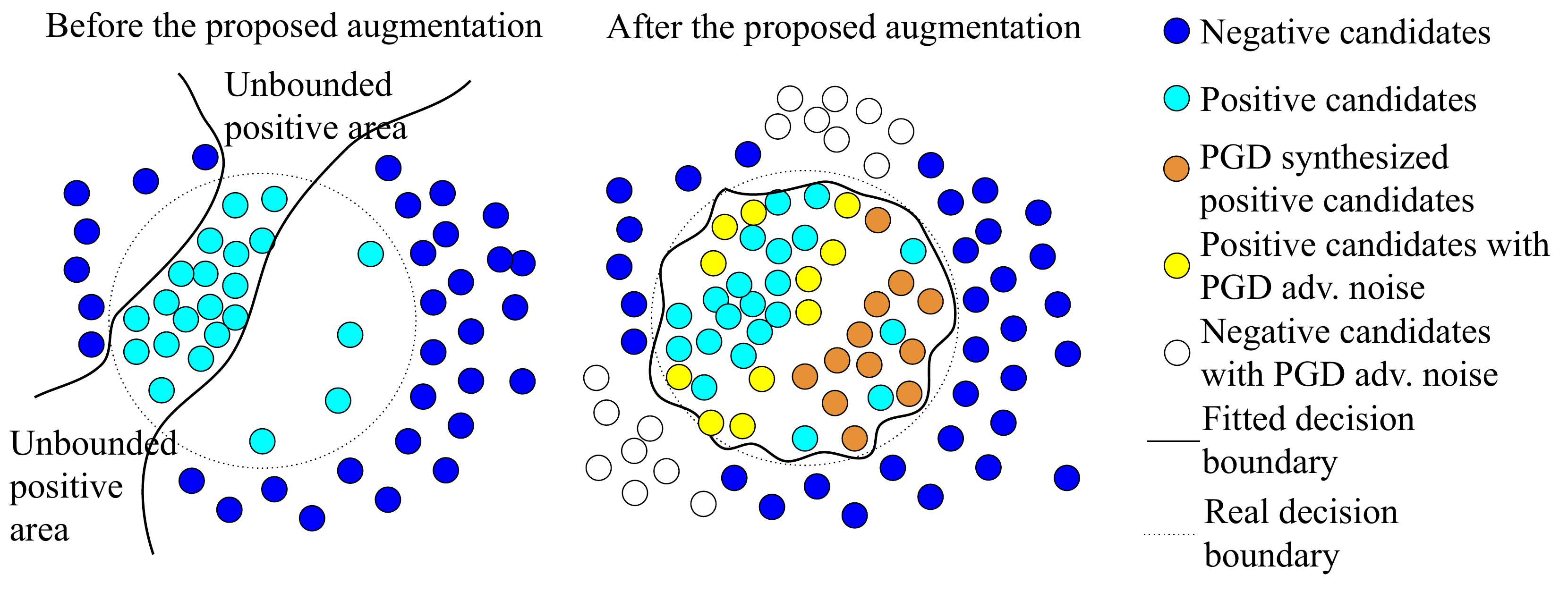}
    \caption{A conceptual illustration of the proposed training scheme. Pulmonary nodules in chest CTs follow a long-tail distribution with rare and hard nodules under-represented. ReLU networks tend to form open decision boundaries which leave the risk for the network to be activated by arbitrary noise \cite{hein2019relu}. We propose adversarial augmentation methods to efficiently search for both hard synthetic nodules and adversarial samples that can improve the robustness of the network.}
    \label{fig:concept}
\end{figure}

Motivated by the LUNA16 challenge \cite{luna16}, many studies have attempted to automate the detection of pulmonary nodules using machine learning, in particular, deep convolutional neural networks (CNN) in order to assist the radiologists in the lung screening workflow \cite{setio2016pulmonary,perez2017automated,dou2016multilevel,zhu2018deeplung,xie2019automated,ding2017accurate}.
Following the coarse-to-fine strategy, the majority of the deep learning-based nodule detection methods are implemented as a two-stage system: (1) a candidate generation network with a large field of view is first trained to output initial detection results with high sensitivity at the cost of low specificity; (2) a false positive reduction (FPR) network is then trained to re-evaluate the confidence of each candidate. 

Though many show CNNs can improve both the sensitivity and the specificity comparing to the previous image processing based CAD systems, CNNs can suffer from a few challenges, which we argue cannot be addressed by simply adding more training data or hyper-parameter tuning.
First, the observer variability among radiologists is known to be high.
For example, only 928 out of 2669 suspected findings from the LIDC-IDRI study are agreed as nodules ($ \geq 3$mm) by all the four radiologists \cite{lidc}. 
Such variability can be caused by factors such as the vague definition of pulmonary nodules, the imbalanced level of expertise among radiologists or the insufficient information provided by chest CT, etc. 
Second, the detection networks tend to miss nodules that are under-represented in the training set, such as the small ground-glass nodules, irregularly shaped nodules or nodules appearing in under-represented contexts.
Because only $3.6\%$ of the screening population have biopsy-proven malignant nodules \cite{NLST},  such malignant nodules can also be under-represented in the training data.
Third, neural networks are known to be prone to unexpected image distortions \cite{easy_to_fool}. Such distortions can happen in the real-world low-dose CT imaging though they are rare in both the training and the benchmark datasets.
As we show later in this paper, even simple noise patterns can determine an under-augmented nodule detector to giving positive responses.
Under- or over-detecting nodules caused by such unanticipated distortions can pose the potential risk of distracting and biasing the radiologists.
Therefore, besides achieving overall high sensitivity and a low number of false positives on clean benchmark datasets, a nodule detection system is also expected to
(1) be capable of detecting under-represented nodules that are rare in both the training and benchmark datasets
(2) be robust to unanticipated noise and distortions in the real-world images.

Motivated by the reasons above, we propose to augment the training set of lung nodule detection by adversarially attacking a pre-trained false positive reduction network with both hard synthetic nodules as well as noise image perturbations.
The concept is illustrated in Fig.~\ref{fig:concept}.
First, we use projected gradient descent (PGD) \cite{madry2017towards} to search for the adversarial samples that can determine a trained false positive reduction network into outputting over-confident wrong predictions.
These searched patches are then added to the training patches to augment the detector to be more robust to both under-represented nodules and unanticipated image distortions.
PGD is used for searching for three types of adversarial augmentation patches:
(1) latent codes to sample hard synthetic nodules that the detector fails to detect;
(2) perturbation noise that can make the nodule detector fail to detect;
(3) noise patterns that can easily determine the nodule detector to giving false-positive findings;
To evaluate the proposed methods, we train a baseline nodule detector following the general 2-stage framework using a large-scale training dataset.
The adversarial patches are then generated by attacking the baseline false positive reduction (FPR) network and are used for augmenting the FPR network.
By evaluating on two different benchmark datasets, we show the proposed techniques can improve the detection performance on clean benchmark data.
Using the same techniques, we also generate adversarial samples to stress-test the trained false positive reduction networks. We show that the augmented networks are more robust to both hard nodules and noise perturbations.

\section{Related Work}
\subsection{Deep learning based nodule detection}
As one of the most popular applications of computer-aided diagnosis systems, many studies have been dedicated to using image processing and machine learning algorithms to detect lung nodules \cite{zhang2018pulmonary}. 
The majority of the nodule detection framework generate candidates first either with an image processing pipeline or a fully convolutional neural network.  
Then a separate classifier is trained to reduce false positives based on the input 3D CT patches centered at the candidate locations.
Most of the recent works were developed based on the LUNA challenge \cite{luna16} which acquired its data from the LIDC-IDRI dataset \cite{lidc}. Though the annotation process of the LIDC-IDRI dataset has been well documented and is considered reliable, the quantity and diversity of the LIDC-IDRI dataset are highly limited. Besides the LUNA challenge, there have been no benchmarks reported with known statistics. Though the metrics computed from the FROC curves are suitable for reporting the detection performance on a given benchmark dataset, it is not often thoroughly investigated that how robust such detection systems would perform on the rare cases as well as noise perturbations.
Our work shows that the conventional CNNs trained without adversarial augmentation would generally fail to recognize rare nodules as well as prone to image noise.
For a more comprehensive review of the deep learning based lung nodule detection systems, we would refer our readers to \cite{pehrson2019automatic,zhang2018pulmonary}.

\subsection{Data synthesis based augmentation in medical image analysis}
Inspired by the recent advances in generative models, there have been increasing interests in synthesizing objects in medical images to augment the existing training set for better diversity \cite{yi2019generative}.
Many recent studies proposed to use generative networks to synthesize lung nodules to improve the performance of diverse lung nodule related applications \cite{yang2019,liu2018decompose,jin2018ct,xu2019tunable,xu2019correlation,gao2019augmenting,han2019synthesizing,wang2019wgan}. Most learning based nodule synthesis methods start with training a generative network to map low dimensional latent codes to realistic lung nodules in chest CT using either variational auto-encoder (VAE) or Generative adversarial networks (GAN).
Latent codes are sampled from a predefined prior distribution randomly to synthesize nodules resembling the real ones. These synthetic nodules are blended into the original image contexts by either formulating the training task as either image inpainting \cite{jin2018ct} or using an extra context-blending network \cite{liu2018decompose}.
In \cite{liu2018decompose}, authors use both the discriminator error and the classification error to select only the hard synthetic cases to be added to the augmented dataset.
We show that such sampling strategies can be inefficient. 
The majority of the synthetic samples would add little values since they can be successfully recognized by a network that is trained on a large-scale dataset. 
However, hard samples can be drawn from a synthesizer without exhaustive search if the latent codes are optimized to increase the training loss of a trained network. In \cite{mayer2020adversarial}, authors showed adversarial sampling can help network generalize better on multi-class classification problems.

\subsection{Over-confident neural networks and adversarial training}
To build robust computer-aided diagnosis systems that are robust to out of distribution (OOD) samples, one can train the network to estimate the decision uncertainty and reject the samples when the estimated uncertainty is high \cite{florin_uncert,evidential_deep}.  
Though we also use the beta distribution in our work for uncertainty estimation \cite{florin_uncert}, we show that the uncertainty estimation techniques alone would be insufficient to make the network robust to avoid over-confident decisions on OOD samples.
In \cite{madry2017towards}, it is argued that ReLU activated neural networks would always have open decision boundaries which leave the risk of high responses for unseen OOD samples. 
In another paper, it is argued that batch normalization is also a cause of the adversarial vulnerability \cite{bn_adv_vul}. 
Such network vulnerability is hard to be reflected by the clean medical image benchmark datasets.
In \cite{meinke2019towards, hein2019relu}, it is proposed to use PGD \cite{madry2017towards} to search for the adversarial augmentation cases from uniform noise or permuted input patches to augment the clean training dataset.
We use similar techniques to adversarially sample both hard positive and hard negative nodule samples to enhance the adversarial robustness of the nodule detection networks. Though it was suggested that the adversarially trained networks can generalize slightly worse on clean data \cite{stutz2019disentangling,2018arXiv180512152T,raghunathan2019adversarial,xie2020intriguing}, we believe such robustness is still vital for real-world medical AI applications.

\subsection{Adversarial robustness of medical image analysis systems}
The vulnerability of CNN against adversarial noise also poses potential risks for deploying the computer-aided diagnosis systems in real clinics as investigated by some recent studies \cite{generalizability_vs_robustness,finlayson2019adversarial,li2019Neurocomputing}.
Some early studies also attempted to defend the networks from adversarial noise using different types of data augmentation, such as using geometric transformation \cite{manifold_exploring} or adding Poisson noise \cite{huang2018some}.
Authors of \cite{mitigating_adversarial} also propose to use the model ensemble to improve the model robustness of nodule malignancy prediction network.
Given the fact that defending against adversarial samples is a challenging task,  \cite{ma2019understanding,li2020robust} also proposed to analyze the neural network feature distributions to detect adversarial samples.

\section{Methods}

\subsection{Baseline Detection Architectures}
\begin{figure}[t]
    \centering
    \includegraphics[width=\linewidth]{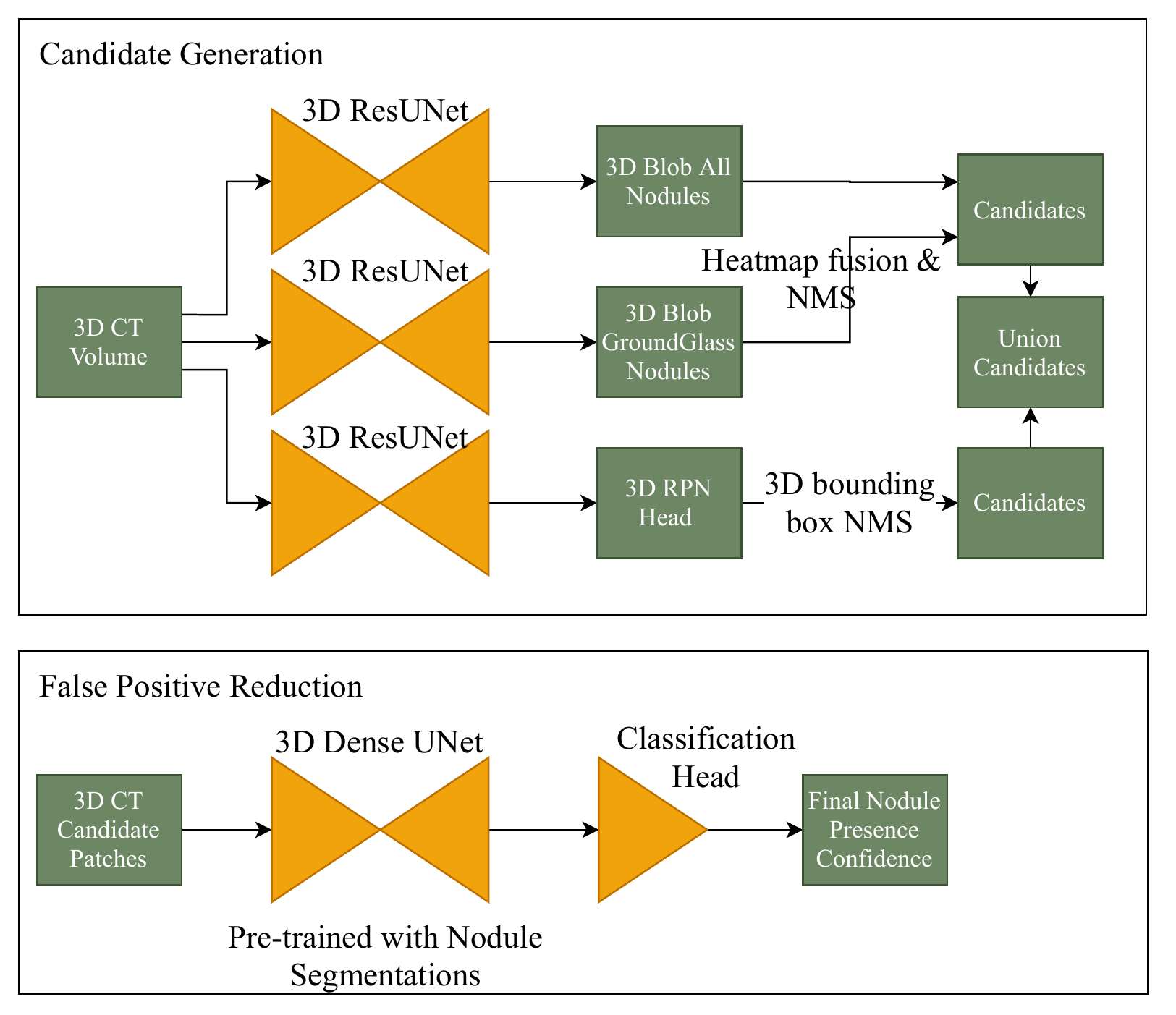}
    \caption{The baseline two stage nodule detection framework used in this work.}
    \label{fig:base}
\end{figure}

Similar to many new deep learning based nodule detection frameworks, our baseline framework consists of a candidate generation (CG) module and a false positive reduction (FPR) module as shown in Fig.~\ref{fig:base}.
The candidate generation module is trained to achieve high sensitivity via over-detecting nodule candidates.
We use three identical 3D ResUNets \cite{resunet} as the CG backbone networks without weight sharing.
The first CG network is first trained to output 3D heatmaps with the nodule centers represented by 3D Gaussian blobs with the same sizes (3D Blob All Nodules). 
We then fine-tune the first CG with only the ground glass candidates and part-solid candidates since they are under-represented in the training set (3D Blob Ground Glass Nodules).
The candidates are derived with non-maximum suppression (NMS) on the fusion heatmap obtained by taking the element-wise maxima of the two network output heatmaps.
We also finetune the first CG network by adding a 3D region proposal network (RPN) head \cite{fasterrcnn} to outputting 3D bounding boxes (3D RPN Head). 
We observed that even though it was hard to improve the sensitivity of the standalone 3D RPN based CG alone (72.00\% and 73.07\% sensitivity in both reported benchmarks), some of the true positive findings are complementary to the blob based CG. By merging the 3D RPN and blob CG candidates, we improved the blob CG sensitivities from 98.00\% and 93.07\% to 100\% and 97.69\% when having 100 candidates per scan.
The final candidates of the system are obtained by taking the union of the blob candidates and the 3D RPN bounding box candidates.

The false-positive reduction module is then trained to re-evaluate the candidates and prune the false-positive findings based on the classification confidence. 
It is built with a DenseUNet network pre-trained with nodule segmentation.
We add shallow classifier layers on top of it to derive the FPR confidence scores.
The network is trained using $64^3$ patches with a resolution of $0.625^3$mm.
We train all the CG and FPR networks using the Adam optimizer \cite{adam} with the initial learning rate $0.001$.

We trained the CG framework first and froze it before performing the analysis presented in this work.
For the brevity of this paper, we demonstrate the proposed techniques only to improve the FPR while assuming the CG networks are trained and frozen. However, the same techniques can also be used for improving CG networks.

\begin{figure}[!htb]
    \centering
    \includegraphics[width=\linewidth]{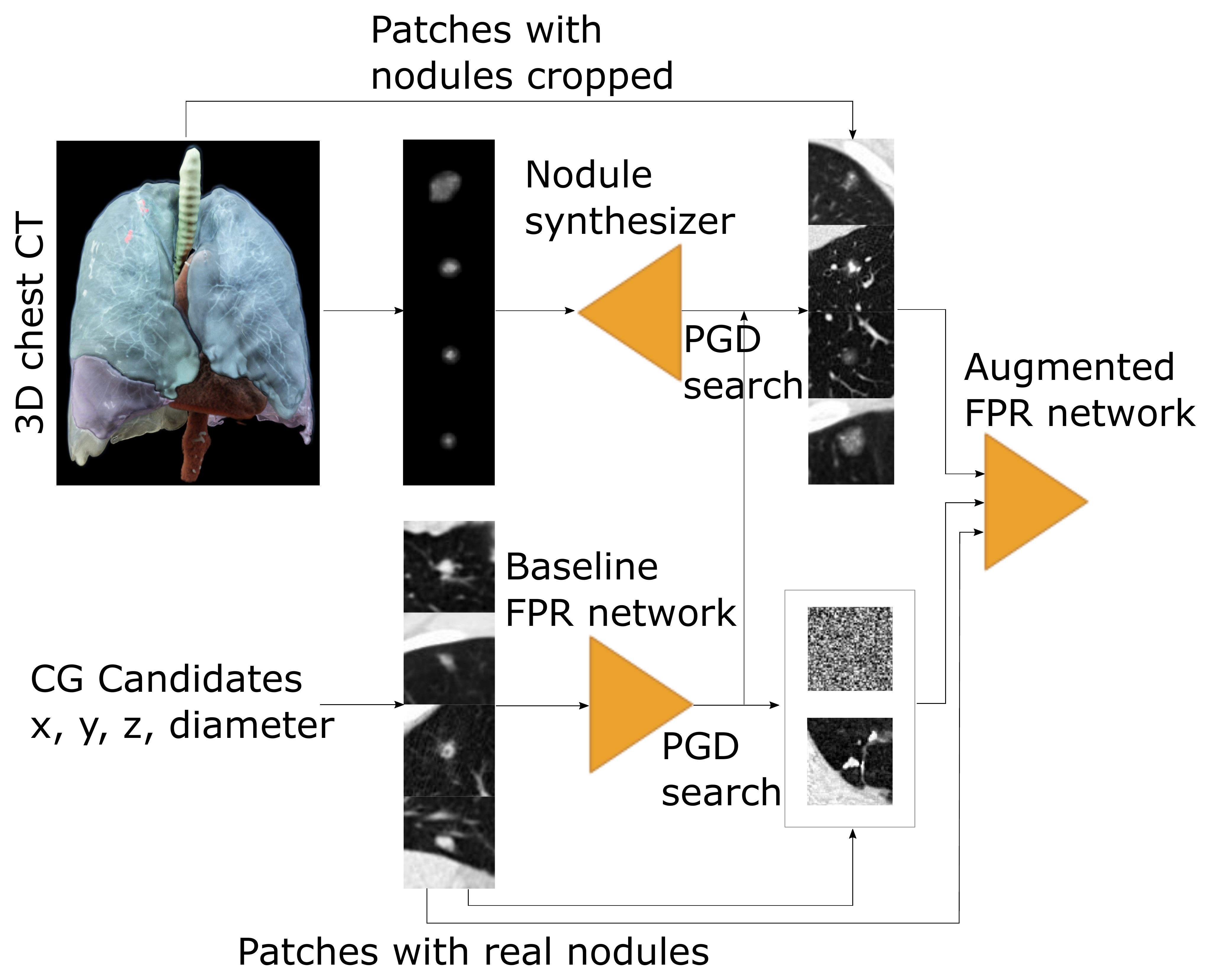}
    \caption{The data-flow illustration of the proposed adversarial augmentation framework for enhancing the false positive reduction (FPR) network in a nodule detection pipeline.}
    \label{fig:dataflow}
\end{figure}

\subsection{Hard-Sample Synthesis with PGD Sampling}
\begin{figure}[!htb]
    \centering
    \includegraphics[width=\linewidth]{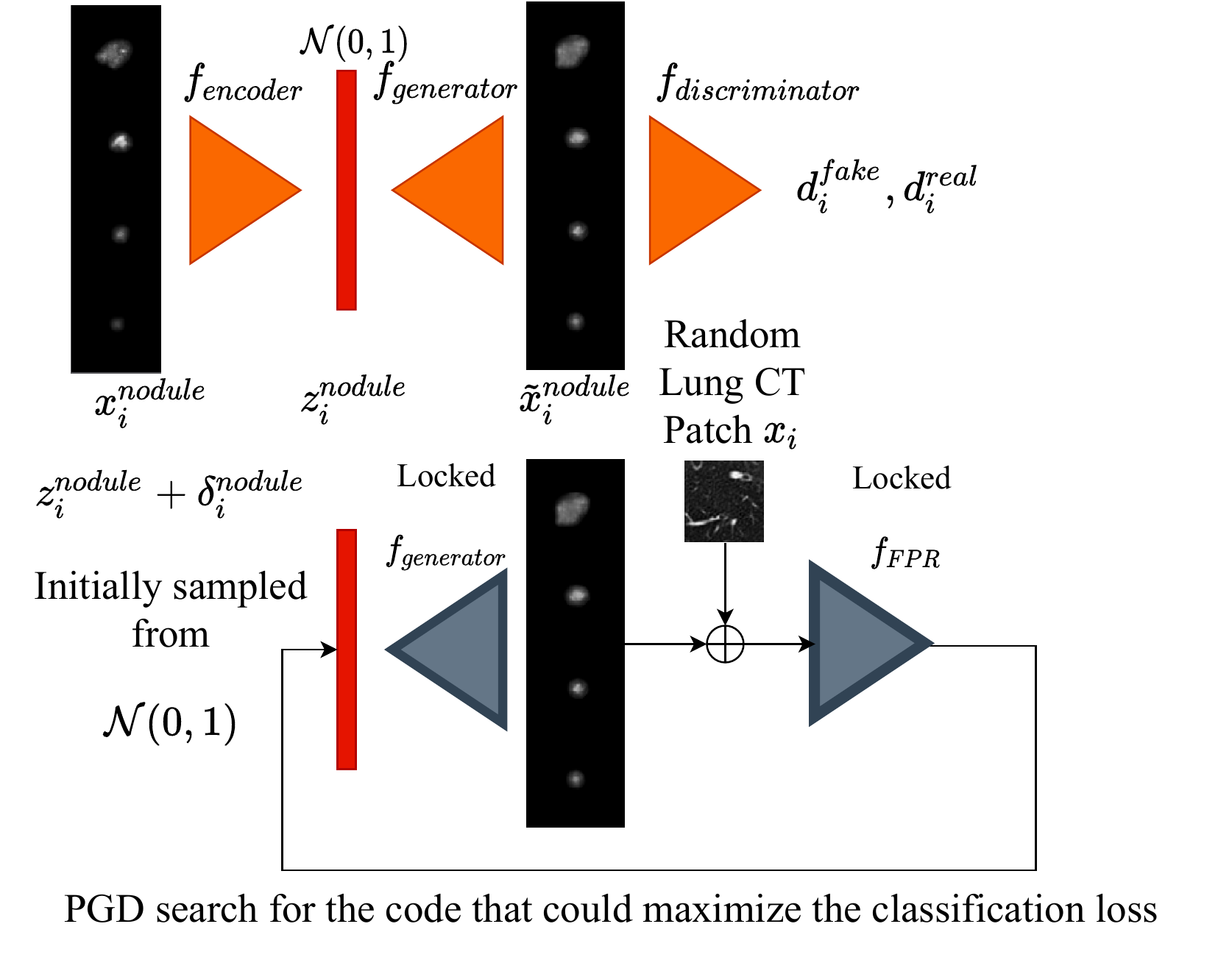}
    \caption{The illustration of the nodule synthesis framework.}
    \label{fig:cvae}
\end{figure}

\begin{figure}[!htb]
    \centering
    \includegraphics[width=\linewidth]{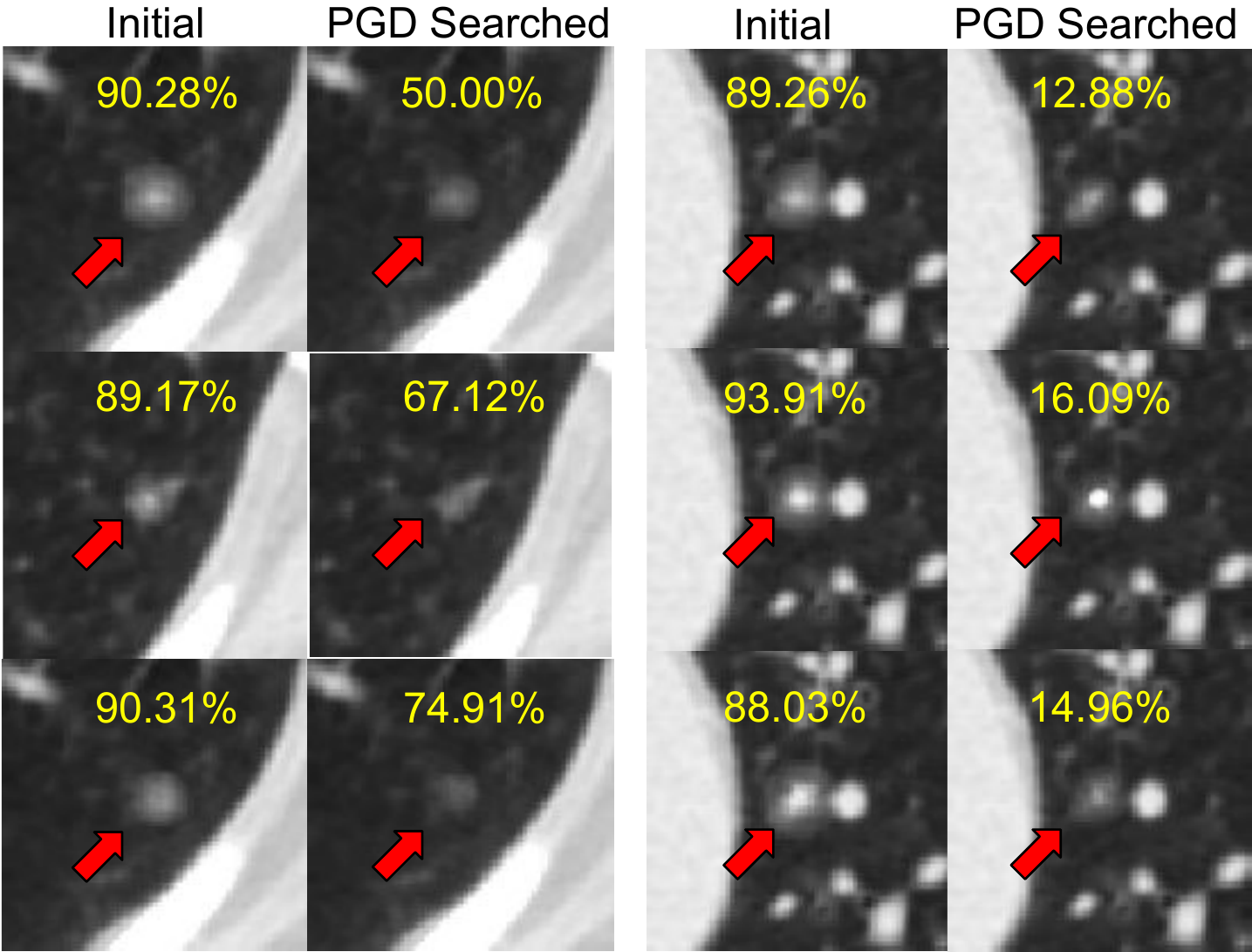}
    \caption{The demonstrations of the synthetic nodules before and after PGD searching. With slight perturbation in the nodule appearance, the nodule detector trained with conventional strategy would output significantly lower confidence score.}
    \label{fig:syn-pgd-search}
\end{figure}

We train a nodule synthesizer $f_{generator}$ that can be controlled by the latent code sampled from a prior distribution.
We implement the $f_{generator}$ with a 3D convolutional variational encoder.
We extract the nodules out of the CT context with the manually annotated nodule segmentation.
The boundary of the nodule segmentation is blurred with a distance transform.
As shown in Fig.~\ref{fig:cvae}, we firstly map the cropped 3D nodules to an encoding space using the encoder network $f_{encoder}$, then the variational encoding is reconstructed back to the nodules in chest CT.
We jointly train a WGAN-GP discriminator \cite{Gulrajani2017ImprovedGANs} with spectral normalization \cite{spectral_norm} to enforce the generator to add high-frequency details to mimic the real nodules in CT. The data flow can be summarized as 
\begin{equation}
\mu_i, \sigma_i = f_{encoder}(x_i^{nodule})
\end{equation}
\begin{equation}
z_i^{nodule} \sim \mathcal{N}(\mu_i, \sigma_i^2)
\end{equation}
\begin{equation}
\tilde{x}_i^{nodule} = f_{generator}(z_i^{nodule})
\end{equation}
\begin{equation}
d_i^{fake}, d_i^{real} = f_{discriminator}(\tilde{x}_i^{nodule}, x_i^{nodule})
\end{equation}
Here $d_i^{fake}$ and $d_i^{real}$ the discriminator output for the fake and real samples.
The training objective of the nodule synthesizer can be summarized as
\begin{equation}
    L_{discriminator} = L_{WGAN-GP}(d_i^{fake}, d_i^{real})
\end{equation}
\begin{equation}
\begin{split}
    L_{encoder} + L_{generator} = |\tilde{x}_i^{nodule} - x_i^{nodule}| + \\ \lambda_1 D_{KL}(\mathcal{N}(\mu_i, \sigma_i ^ 2) || \mathcal{N}(0, 1)) - \\
    \lambda_2 L_{WGAN-GP}(d_i^{fake}, d_i^{real})
\end{split}
\end{equation}
where $D_{KL}(\mathcal{N}(\mu_i, \sigma_i ^ 2) || \mathcal{N}(0, 1))$ optimizes the probability distribution parameters $\mu$ and $\theta$ to closely resemble that of $\mathcal{N}(0,1)$. $\lambda_2 L_{WGAN-GP}(d_i^{fake}, d_i^{real})$ is the wasserstein GAN discriminator loss regularized by the gradient penalty defined in \cite{Gulrajani2017ImprovedGANs}. In our experiments, we set $\lambda_1 = 10^{-5}$ and $\lambda_2 = 0.1$.

Once the synthesizer is trained, we discard both the encoder network and the discriminator. Only the generator network is kept for sampling synthetic nodules. 
Random nodules can be sampled by feeding a code to the trained generator $f_{generator}(z_i^{nodule} \sim \mathcal{N}(\mu_i, \sigma_i^2))$.
The synthesized nodule can be fused to a random background chest CT patch $x_i$ and then fed to a trained FPR classifier $f_{FPR}$. For the rest of the paper, we define the differentiable fusion of synthetic nodule and the background as $\tilde{x}_i^{nodule} \oplus x_i = \tilde{x}_i^{nodule} * m_i + x_i * (1 - m_i)$ where $m_i$ is a binary mask obtained by thresholding $\tilde{x}_i^{nodule}$.
Though it is feasible to add another training stage as described in \cite{liu2018decompose} to further blend the generated nodule into its context, we found it non-critical for the sake of improving the nodule detection in practice. 
It is inefficient to draw hard-cases directly by randomly sampling from the prior because most of the cases close to the mean have already been learned by the nodule false positive reduction network $f_{FPR}$. 
So instead of randomly sampling the encoding of nodules, we use the projected gradient descent (PGD) as originally used for generating adversarial attacks \cite{madry2017towards} to sample hard nodules.
For each sampling, we initialize the encoding from the standard normal distribution $z_i^{nodule} \sim \mathcal{N}(0,1)$ and randomly initialize a perturbation vector $\delta_i^{nodule}$ to explore the neighbourhood $\mathcal{S}$ of $z_i^{nodule}$ within a bounded radius.
$\delta_i^{nodule}$ is updated by PGD to maximize the $L_{FPR}$ as
\begin{equation}
    e_i = f_{FPR}( f_{generator}(z_i^{nodule} + \delta_i^{nodule}) \oplus x_i)
\end{equation}
\begin{equation}
    arg \max_{\| \delta \| \leq \epsilon} {L_{FPR} (e_i, 1)}
\end{equation}
Here, $\oplus$ is the fusion operator that blends the synthetic nodule into the CT context patch $x_i$. 
As an alternative to the combination of the sigmoid activation and the binary cross-entropy loss, We also use the beta distribution as in \cite{florin_uncert} to measure the classification uncertainty in our experiments.
With the beta distribution output, the FPR network $f_{FPR}$ outputs the classification evidence $e_i$ for positive and negative labels.
$L_{FPR} (e_i, 1)$ is the classification loss defined with the beta distribution distance.
For the symbolic brevity, we refer to \cite{florin_uncert} for the detailed definition of the beta distribution network output and the loss function.
It was shown in \cite{florin_uncert} that the beta distribution networks are less likely to be activated by out-of-distribution (OOD) patches and can produce comparable classification accuracy as cross-entropy.
However, in our experiments, we also show that uncertainty estimation alone does not suffice to make the network robust to all OOD samples especially when such samples are searched by PGD. 
The perturbation vector $\delta$ can be updated as 
\begin{equation}
    \delta := \mathcal{P}(\delta + \alpha \nabla_{\delta} L_{FPR}(.))
\end{equation}
where $\mathcal{P}$ denotes the projection onto the ball of interest defined by $\epsilon$; $\alpha$ is the step size.
In our experiments, we set $\epsilon = 0.15$, $\alpha=0.05$. $\delta$ is updated with 20 iterations for each search.
In Fig.~\ref{fig:syn-pgd-search}, we show initial synthetic nodules together with the synthetic nodules searched with PGD. Though visually similar, the tiny differences in the nodule appearance can result in a large difference in the $F_{FPR}$ responses.

\begin{figure}[!htb]
    \centering
    \includegraphics[width=\linewidth]{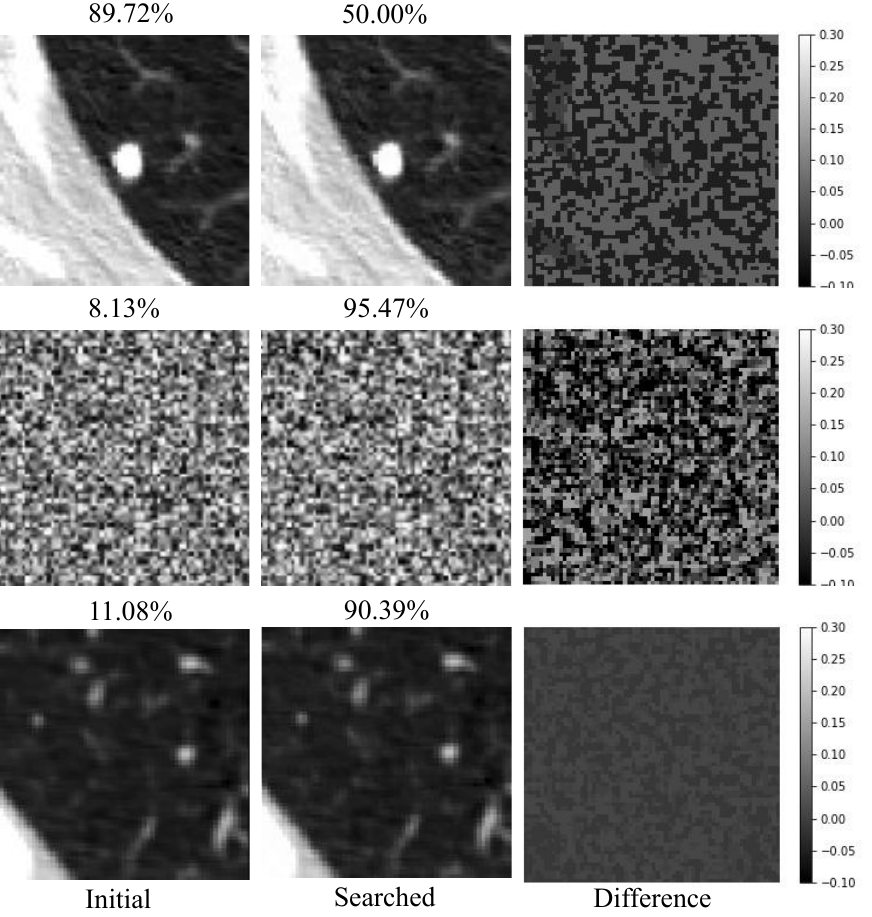}
    \caption{The upper row demonstrates the noise perturbation on nodule patches. Arbitrary noise can determine a trained nodule detector to ignore a well-defined nodule. The middle and bottom row demonstrates that specific noise patterns can activate a trained nodule detector to output high confidence scores from either pure adversarial noise or the negative CT patches distorted by adversarial noise. The difference patches are shown with the window $[0, 0.3]$ to make the perturbation visible while the image patches are shown with the window $[0, 1]$.}
    \label{fig:adv_attack}
\end{figure}

\subsection{Over-confident Perturbation with PGD Sampling}
Besides searching the latent codes for the nodule synthesizer, PGD can also be used for perturbing the real patches $x_i$ as
\begin{equation}
    e_i = f_{FPR}(x_i + \delta_i^{patch})
\end{equation}
\begin{equation} \label{eq:pos_adv_search}
    arg \max_{\| \delta^{patch}_i \| \leq \epsilon} {L_{FPR} (e_i, g_i)}
\end{equation}
where $g_i$ is the groundtruth label for patch $x_i$.
As shown in the first row of Fig.~\ref{fig:adv_attack}, we found for most of the positive nodules patches, it is easy to find a $\delta_i^{patch}$ with a small magnitude to perturb $x_i$ so that $f_{FPR}$ no longer recognizes the nodule resides in it. 
Such perturbations can disturb the model from recognizing the nodules when the images contain unexpected abnormalities, strong imaging artefacts or malicious noise injections.

We also found that even for noise patches $x_i^{uniform}$ drawn from a uniform distribution,
PGD can search for a neighbouring patch and excites the FPR network to output a positive decision, though the searched patch does not contain any interpretable patterns as shown in the second row of Fig.~\ref{fig:adv_attack}.
The intersection between the chest CT distribution and the uniform distribution is expected to have close to zero probability mass.
As explained in \cite{hein2019relu}, ReLU networks decompose the observation space into a finite set of polytopes in which outer polytopes extend to infinity.
Adding the adversarial patches searched by Eq.(\ref{eq:pos_adv_search}) to augment the FPR network can make it robust to such image perturbations by closing the decision boundary.

In practice, we train a baseline FPR network first by randomly sampling real positive and negative candidate patches with $50\%$ chance each until reaching convergence. Then we finetune the baseline model by also sampling from the augmentation patches generated by attacking the baseline model.
For positive sampling, we draw 50\% from the real positive patches, and $25\%$ from synthetic nodules and $25\%$ the adversarial positive patches. For negative sampling, we draw $50\%$ from both real negative patches and $50\%$ from the adversarial negative patches.

\section{Data and Experiment Settings}
6488 3D chest CT scans were collected for training. 
The training images were collected from multiple sources, including the LUNA challenge \cite{luna16}, the NLST cohort \cite{NLST} and an in-house data collection.
Each training image contains at least one radiologist confirmed nodule.
We annotated the nodule locations and diameters in the training images from our in-house dataset and the NLST subset.
Our annotators firstly detected all the potential nodule candidates.
Then two radiologists went through all the candidates to confirm the presence of a nodule. 
$10\%$ of the training images were randomly sampled as the validation set for parameter searching and early stopping.
To evaluate the performance, we constructed two benchmark datasets, as summarized in Table~\ref{tab:data}.
The In-house Benchmark was built based on a private data collection with 174 challenging CT scans.
Besides lung nodules, many patients in the In-house Benchmark also had other types of pulmonary abnormalities which constitute a significant source of false positives for both human readers and the networks.
The NLST Benchmark consists of randomly sampled 272 baseline CT scans from the NLST cohort. The patients were sampled following the real-world screening distribution \cite{NLST} ($1\%$ with cancer, $25.8\%$ with cancer negative nodules and $73.2\%$ healthy) while ensuring (1) the slice thicknesses are lower than $1.5mm$ (2) there is no gap in the DICOM series (3) each image contains the entire lung. 
We had three on-board radiologists read the images in both benchmark datasets independently. 
In the first round, each radiologist marked the nodule candidates individually.
All the candidate nodules spotted in the first round were merged and presented to each radiologist to confirm in case there were under-attended nodule candidates. We took the nodules that are the consensus among all three radiologists as the positive locations while the rest as irrelevant findings which were not involved in the metrics computing. 
We only considered the nodules with the diameters larger than $6mm$ for benchmarking.
However, we do not claim this is a critical choice since the size threshold can be adjusted according to the different application scenarios.

All the augmentation patches, including the synthetic nodules, perturbed positive nodule patches and the perturbation noises, were pre-computed and randomly sampled during the FPR model training by attacking the baseline network (baseline-beta-finetune). Therefore, the large pool of randomly generated patches was kept consistent across different experiments to guarantee the reproducibility of all the experiment results.
We generated synthetic nodule patches on 10 random background patches from each training image.
The locations of the background patches were constrained within the lungs using the lung segmentation masks predicted by a previously trained network. 
We also ensured that the background patches do not contain a real nodule inside.
For each background patch, we sampled the synthesizer six times with random sampling and the PGD sampling, respectively.
It resulted in 389,280 synthetic nodule patches for both sampling strategies.
We generated one adversarially perturbed patch for each positive nodule candidate in our training data (22,169 relevant nodules) similarly to the upper row of Fig.~\ref{fig:adv_attack}. We also generated 100,000 pure adversarial noise patches similarly to the lower row of Fig.~\ref{fig:adv_attack}.
To stress-test the robustness of network at random pulmonary locations, we sampled 10 random patches centered in the lungs as the negative stress-test samples from each benchmark CT volume, while avoiding annotated nodules. We add adversarial noise to these negative samples by attacking the baseline network (baseline-beta-finetune). For all the experiments involving Poisson noise, we used the same ratio to sample the Poisson noise injected patches as used for the adversarial noise patches.

\begin{table}[]
\centering
\caption{The summary of the chest CT benchmark datasets.}
\label{tab:data}
\begin{tabular}{@{}l|ll@{}}
                           & In-house Benchmark & NLST Benchmark \\ \midrule
CT scans                     & 174         & 272         \\
CT scans w/ Nodules          & 97          & 83          \\ \midrule
Solid Nodules              & 94          & 103         \\
Fully Calcified Nodules    & 7           & 19          \\
Part-Solid Nodules         & 13          & 3           \\
Ground Glass Nodules       & 36          & 6           \\
Total Nodules (\textgreater{}=6mm) & 150         & 131         \\ \bottomrule
\end{tabular}
\end{table}

\section{Results}

\begin{figure}[!htb]
     \centering
     \begin{subfigure}[b]{0.45\linewidth}
         \centering
         \includegraphics[width=\textwidth]{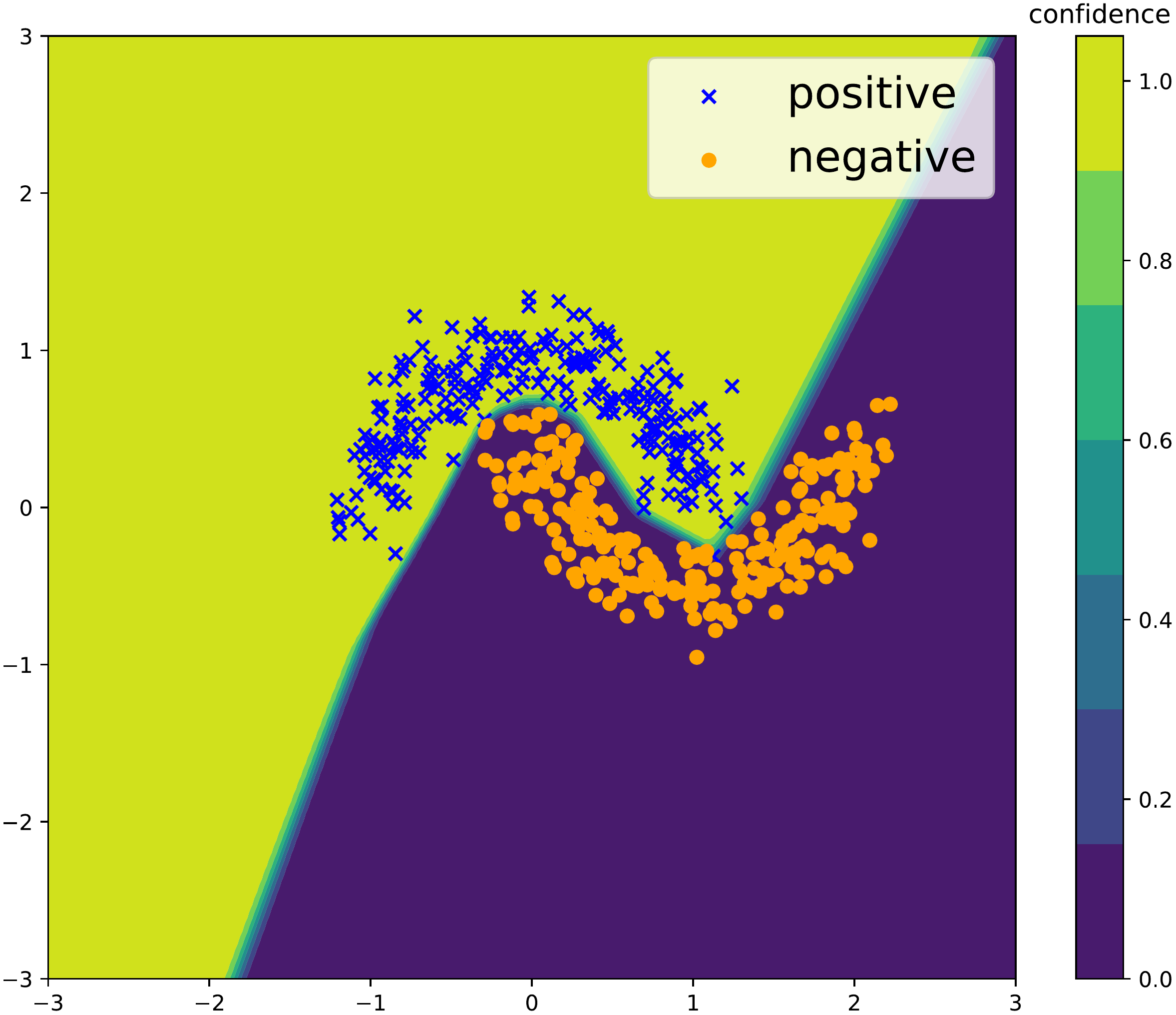}
         \caption{Fully sampled points}
         \label{fig:toy_full}
     \end{subfigure}
     \begin{subfigure}[b]{0.45\linewidth}
         \centering
         \includegraphics[width=\textwidth]{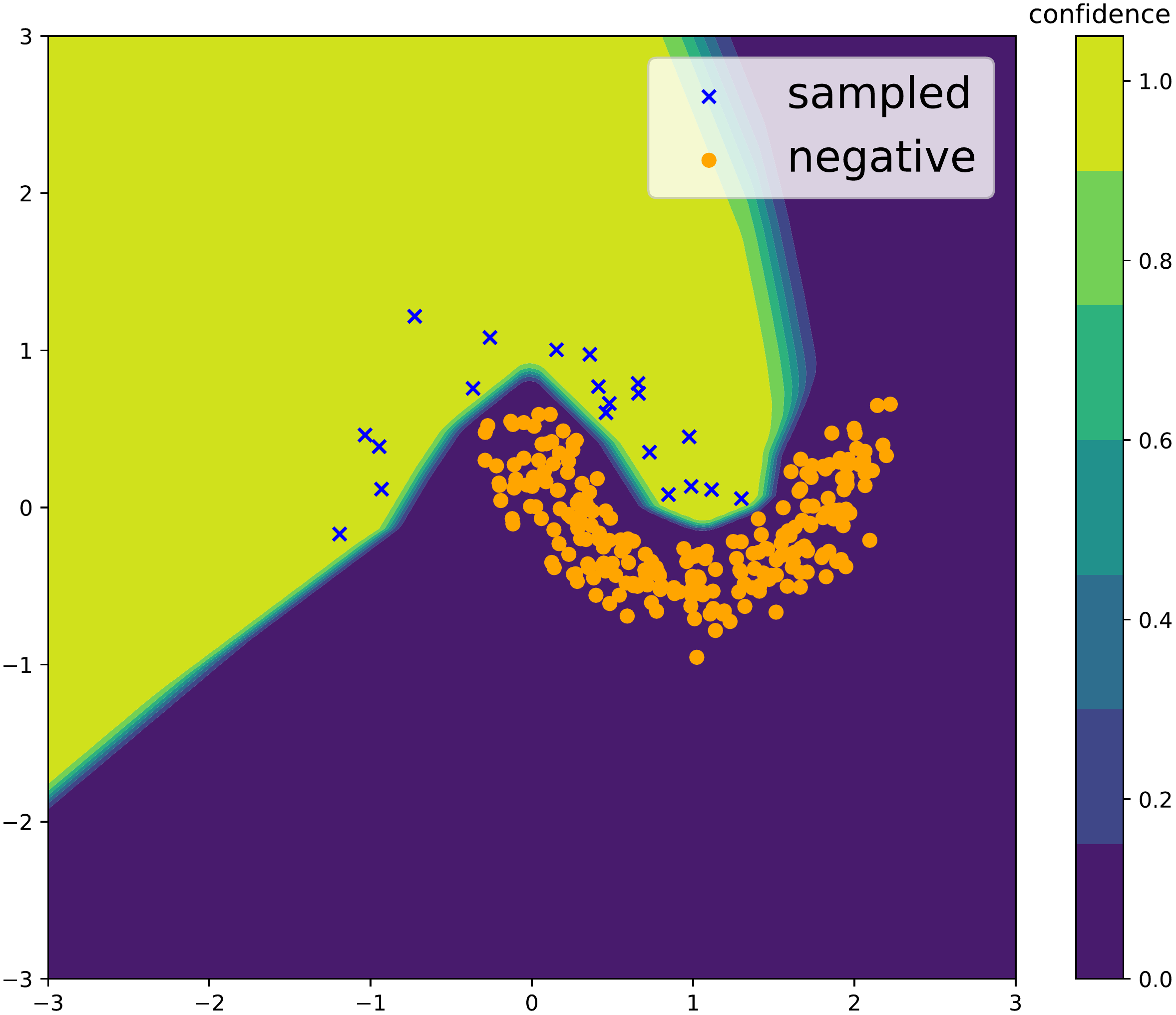}
         \caption{Under sampled points}
         \label{fig:toy_sampled}
     \end{subfigure}
     \begin{subfigure}[b]{0.45\linewidth}
         \centering
         \includegraphics[width=\textwidth]{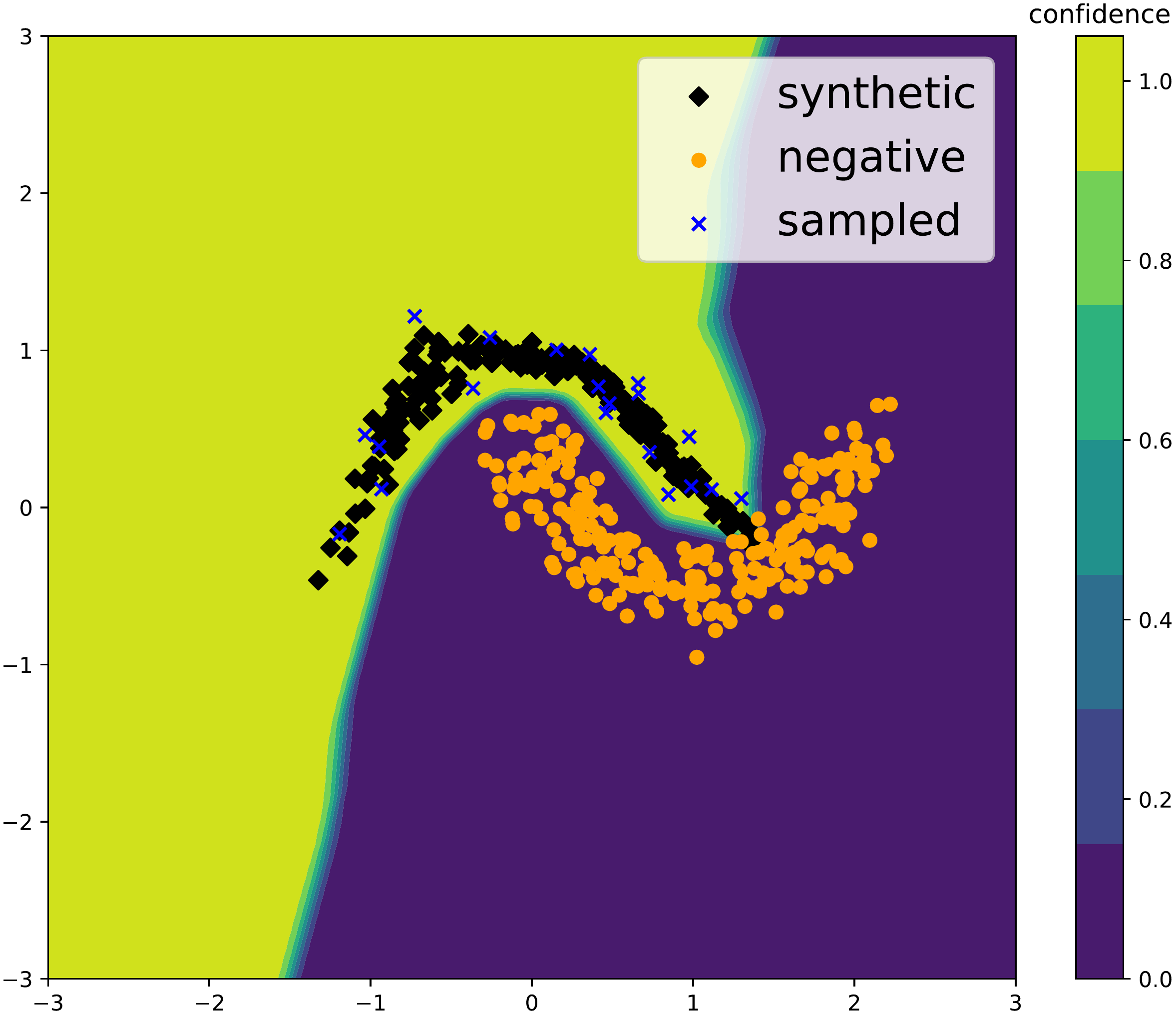}
         \caption{Add synthetic points}
         \label{fig:toy_syn}
     \end{subfigure}
     \begin{subfigure}[b]{0.45\linewidth}
         \centering
         \includegraphics[width=\textwidth]{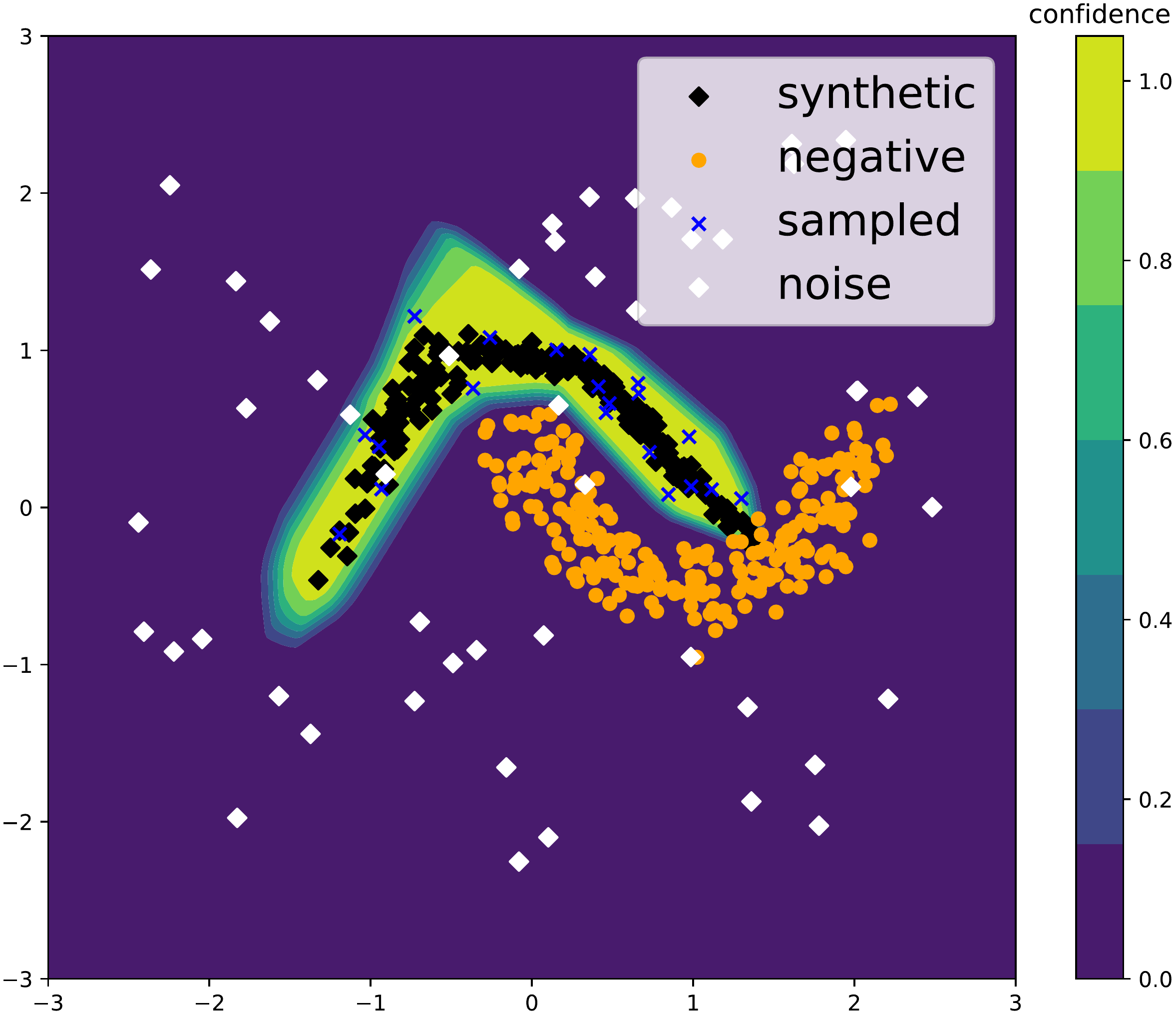}
         \caption{Add uniform noise points}
         \label{fig:toy_syn_noise}
     \end{subfigure}
     \begin{subfigure}[b]{0.45\linewidth}
         \centering
         \includegraphics[width=\textwidth]{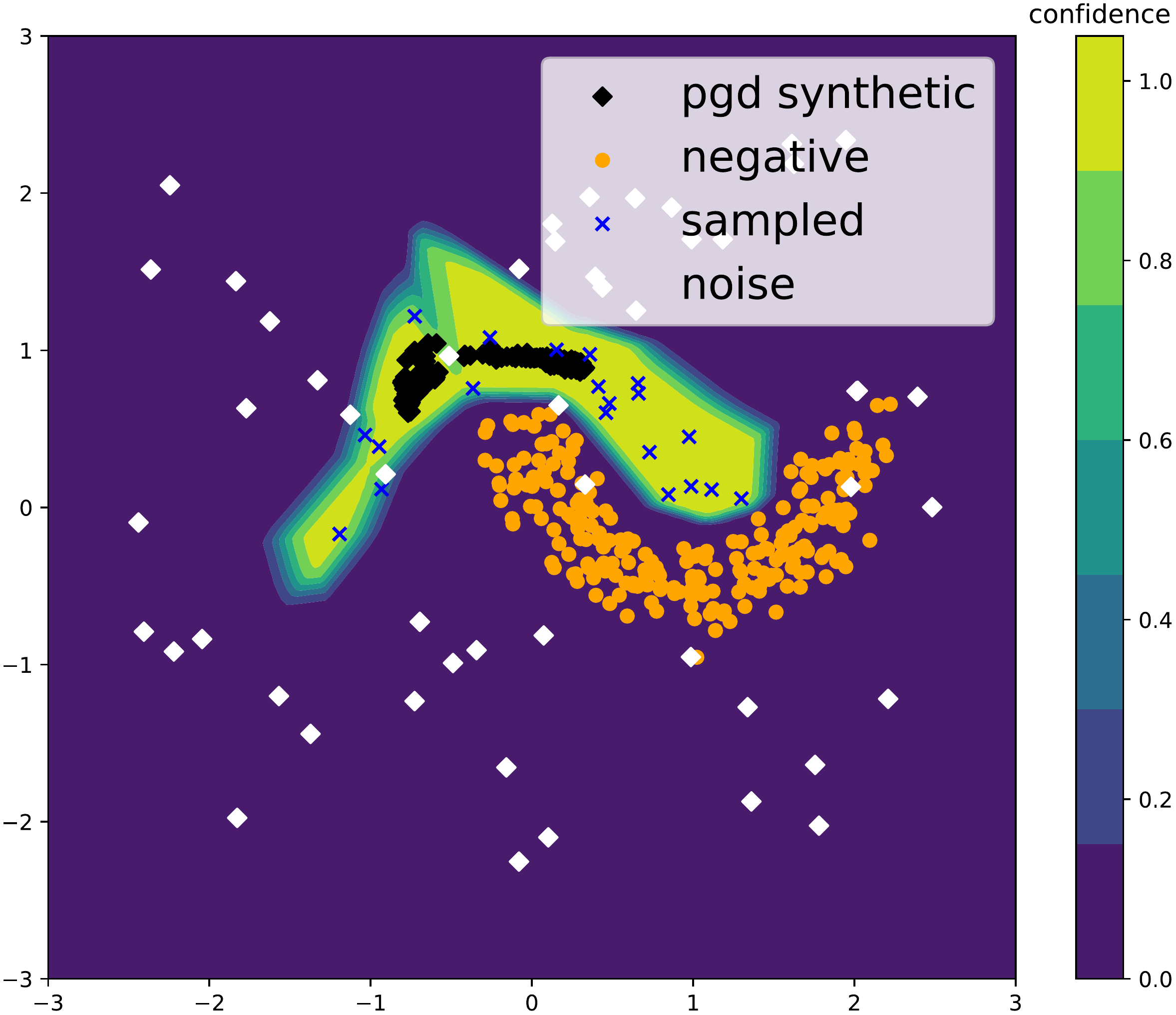}
         \caption{Add PGD synthetic points}
         \label{fig:toy_pgdsyn_noise}
     \end{subfigure}
     \begin{subfigure}[b]{0.45\linewidth}
         \centering
         \includegraphics[width=\textwidth]{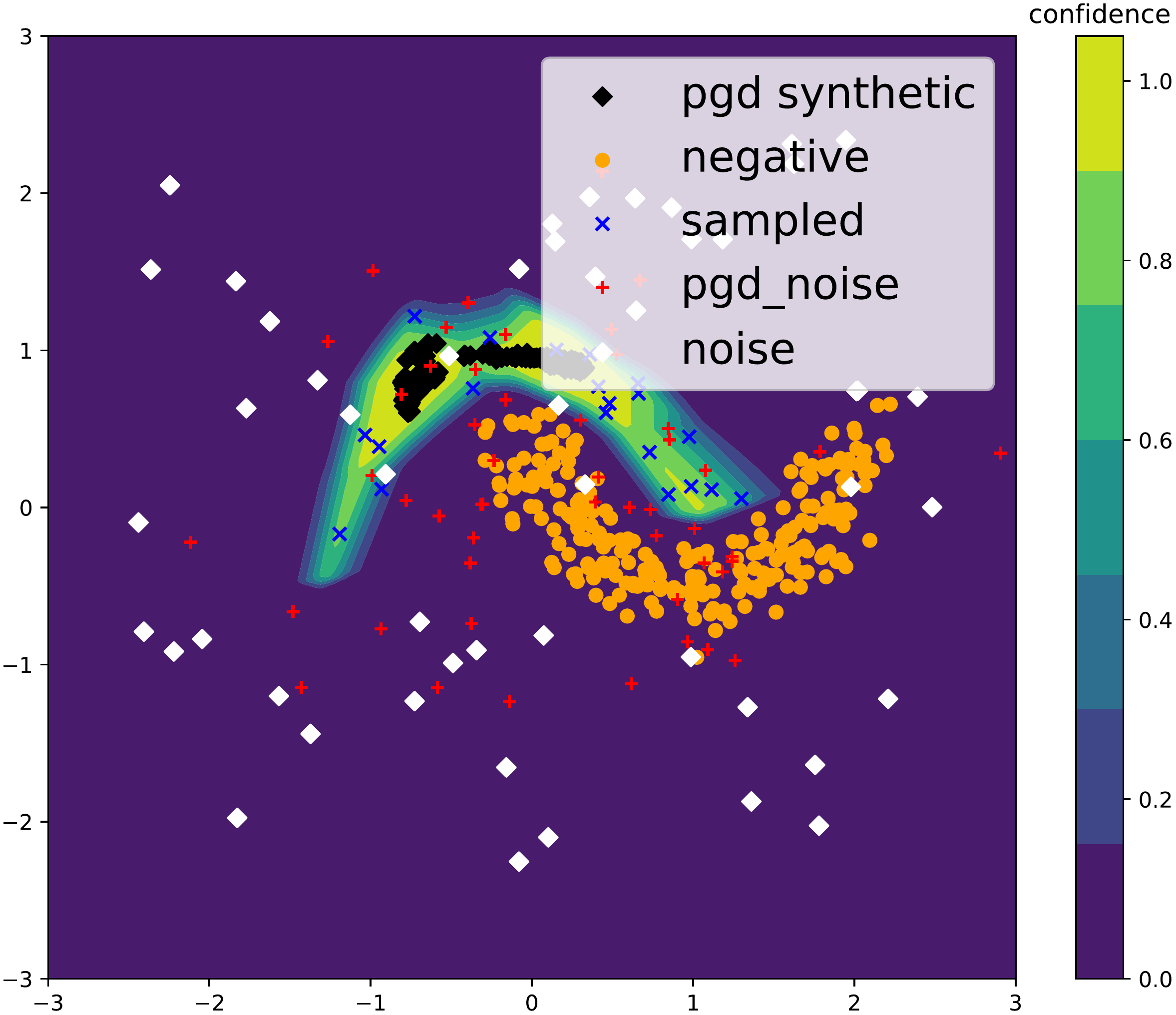}
         \caption{Add PGD noise points}
         \label{fig:toy_pgdsyn_pgdnoise}
     \end{subfigure}
\caption{A toy experiment to depict the concept of the proposed augmentation methods.}
\label{fig:toy}
\end{figure}

\subsection{Toy Example}
In Fig.~\ref{fig:toy}, we firstly show a toy experiment built with the simple two-moon dataset to demonstrate the presented concept.
500 spots are sampled from both the positive and the negative cluster by adding the Gaussian noise with the standard deviation of $0.15$. In our context, they represent the positive and negative candidates used for training the FPR classifier.
We train a ReLU activated multi-layer perceptron to mimic the FPR classifier based on the sampled spots to plot the decision boundary.
We then sub-sample only 20 positive candidates following a long tail distribution to simulate the real-world training set distribution as Fig.~\ref{fig:toy_sampled}.
We trained a small VAE on the 20 positive spots and generated synthetic samples by drawing the latent code from a standard normal distribution.
The added synthetic spots help filling the hole in the decision boundary as in Fig.~\ref{fig:toy_syn}. 
However, a sizeable out-of-distribution area is also predicted as confident positive as anticipated in \cite{hein2019relu}.
We then sampled another 20 spots that are randomly drawn from a uniform distribution and added them to the negative cluster.
In Fig.~\ref{fig:toy_syn_noise}, it is shown that such noise samples can bound the decision boundary tightly to the positive cluster.
Though there is a small chance that the noise spots can also reside in the positive cluster, such cases are extremely rare in the real world 3D inputs.
Though we use uniform sampling in this toy example, it is notable that in a high-dimensional input space, the random sampling can be highly in-efficient for both synthesizing real nodules and generating adversarial noise samples.
We use PGD to search for the latent code from the trained VAE. As in Fig.~\ref{fig:toy_pgdsyn_noise}, the PGD searched synthetic spots only reside in the under-sampled region. In addition to the uniform spots, we show the PGD searched negative spots which are closer to the positive cluster in Fig.~\ref{fig:toy_pgdsyn_pgdnoise}.
Such supporting negative spots can be more efficient for refining the decision boundary when the input dimension is higher as in 3D chest CT patches.

\subsection{Benchmark on clean data}
\begin{table}[]
\centering
\caption{The table summarizes the complexities of the nodule detection models.}
\label{tab:complexity}
\begin{tabular}{@{}llll@{}}
\toprule
Network                  & \#Param & Mac (G) & Input Size  \\ \midrule
DeepLung \cite{zhu2018deeplung} & 5.36M   & 168.43  & 128$\times$128$\times$128 \\
Blob Solid / GGO ResUNet & 143,92k & 67.15   & 128$\times$128$\times$128 \\
RPN                      & 276.93k & 113.91  & 129$\times$128$\times$128 \\
FPR                      & 455.98k & 17.24   & 64$\times$64$\times$64    \\ \bottomrule
\end{tabular}
\end{table}

\begin{table*}[]
\caption{The table summarizes the FROC metrics on the clean benchmark datasets. The CPM score averages the sensitivities sampled at 7 log-scale operating points indicating different numbers of false positives  (0.125, 0.25, 0.5, 1, 2, 4, 8).}
\label{tab:froc}
\begin{adjustbox}{width=\textwidth,center}
\begin{tabular}{@{}llllllllllll@{}}
\toprule
\multicolumn{12}{l}{In-house Benchmark}\\ \midrule
\multicolumn{1}{l|}{}                       & PERTURB & SYN & \multicolumn{1}{l|}{LOSS} & \bf{CPM}     & FP=0.125   & FP=0.25    & FP=0.5     & FP=1       & FP=2       & FP=4       & FP=8       \\ \midrule
\multicolumn{1}{l|}{baseline-DeepLung \cite{zhu2018deeplung}}            & \xmark       & \xmark   & \multicolumn{1}{l|}{N/A}   & 55.15\%	(46.56\% - 63.92\%)
 & 31.59\% & 38.81\% & 51.84\% & 58.64\% & 62.98\% & 68.59\% & 73.58\% \\\midrule
\multicolumn{1}{l|}{baseline-ce}            & \xmark       & \xmark   & \multicolumn{1}{l|}{CE}   & 88.46\% (83.18\% - 93.76\%) & 73.09\% & 82.24\% & 89.84\% & 92.02\% & 92.28\% & 94.65\% & 96.64\% \\
\multicolumn{1}{l|}{baseline-beta}          & \xmark       & \xmark   & \multicolumn{1}{l|}{BETA} & 89.11\% (82.91\% - 93.87\%) & 75.43\% & 83.93\% & 88.78\% & 90.80\% & 91.77\% & 94.26\% & 96.64\% \\
\multicolumn{1}{l|}{baseline-beta-finetune} & \xmark       & \xmark   & \multicolumn{1}{l|}{BETA} & 88.90\% (83.62\% - 94.22\%) & 75.58\% & 84.29\% & 90.79\% & 91.37\% & 92.11\% & 94.24\% & 96.70\% \\ \midrule
\multicolumn{1}{l|}{beta+syn (random)}      & \xmark       & \cmark   & \multicolumn{1}{l|}{BETA} & 90.76\% (85.95\% - 95.49\%) & 79.89\% & 87.43\% & 92.05\% & 93.34\% & 93.35\% & 94.26\% & 96.67\% \\
\multicolumn{1}{l|}{beta+syn}               & \xmark       & \cmark   & \multicolumn{1}{l|}{BETA} & 91.22\% (86.13\% - 95.69\%) & 81.09\% & 87.66\% & \bf{92.66\%} & 93.33\% & 93.33\% & 94.17\% & 96.99\% \\
\multicolumn{1}{l|}{beta+poisson}           & \xmark       & \xmark   & \multicolumn{1}{l|}{BETA} & 91.98\% (86.74\% - 96.30\%) & 81.50\% & 87.24\% & 90.62\% & 93.40\% & \bf{95.87\%} & \bf{97.29\%} & 97.96\%  \\ 
\multicolumn{1}{l|}{beta+poisson+syn}       & \xmark       & \cmark   & \multicolumn{1}{l|}{BETA} & \bf{92.26\%} (87.52\% - 96.30\%) & \bf{81.96\%} & \bf{88.16\%} & 92.65\% & \bf{94.00\%} & 95.10\% & 96.04\% & \bf{98.00\%} \\
\multicolumn{1}{l|}{beta+perturb}           & \cmark       & \xmark   & \multicolumn{1}{l|}{BETA} & 90.07\% (85.19\% - 95.02\%) & 76.35\% & 85.83\% & 89.91\% & 93.42\% & 93.61\% & 95.40\% & 97.92\% \\
\multicolumn{1}{l|}{beta+perturb+syn}       & \cmark       & \cmark   & \multicolumn{1}{l|}{BETA} & 90.47\% (85.22 - 95.12) & 77.52\% & 86.29\% & 89.95\% & 92.75\% & 93.97\% & 94.98\% & 97.45\% \\ \midrule
\multicolumn{12}{l}{NLST Benchmark}                                                                                                                                        \\ \midrule
\multicolumn{1}{l|}{}                       & PERTURB & SYN & \multicolumn{1}{l|}{LOSS} & \bf{CPM}     & FP=0.125 & FP=0.25 & FP=0.5 & FP=1  & FP=2       & FP=4       & FP=8       \\ \midrule
\multicolumn{1}{l|}{baseline-DeepLung \cite{zhu2018deeplung}}            & \xmark       & \xmark   & \multicolumn{1}{l|}{N/A}   & 75.71\%	(67.16\% - 84.32\%)
 & 32.77\% & 45.05\% & 52.14\% & 57.07\% & 67.21\% & 74.09\% & 75.71\% \\\midrule
\multicolumn{1}{l|}{baseline-ce}            & \xmark       & \xmark   & \multicolumn{1}{l|}{CE}   & 82.56\% (73.26\% - 91.03\%) & 52.18\% & 71.50\% & 84.99\% & 89.68\% & 91.68\% & 93.14\% & 95.63\% \\
\multicolumn{1}{l|}{baseline-beta}          & \xmark       & \xmark   & \multicolumn{1}{l|}{BETA} & 80.62\% (69.49\% - 89.78\%) & 44.74\% & 68.39\% & 83.35\% & 88.56\% & 91.38\% & 93.18\% & 94.40\% \\
\multicolumn{1}{l|}{baseline-beta-finetune} & \xmark       & \xmark   & \multicolumn{1}{l|}{BETA} & 83.60\% (74.19\% - 91.91\%) & 53.69\% & 73.44\% & 85.30\% & 91.35\% & 93.01\% & 93.99\% & \bf{95.71\%} \\ \midrule
\multicolumn{1}{l|}{beta+syn (random)}      & \xmark       & \cmark   & \multicolumn{1}{l|}{BETA} & 85.81\% (77.40\% - 93.43\%) & 66.04\% & 80.10\% & 86.55\% & 90.17\% & 92.05\% & 93.15\% & 95.06\% \\
\multicolumn{1}{l|}{beta+syn}               & \xmark       & \cmark   & \multicolumn{1}{l|}{BETA} & \bf{87.89\%} (81.20\% - 93.93\%) & \bf{74.44\%} & \bf{81.04\%} & 87.61\% & \bf{90.55\%} & 93.30\% & 93.80\% & 94.77\% \\
\multicolumn{1}{l|}{beta+poisson}           & \xmark       & \xmark   & \multicolumn{1}{l|}{BETA} & 86.33\% (76.71\% - 93.46\%) & 64.95\% & 78.22\% & 87.41\% & 90.25\% & \bf{93.85\%} & \bf{94.18\%} & 95.44\%  \\ 
\multicolumn{1}{l|}{beta+poisson+syn}       & \xmark       & \cmark   & \multicolumn{1}{l|}{BETA} & 86.55\% (78.16\% - 93.52\%) & 68.25\% & 78.33\% & 86.96\% & 90.46\% & 93.14\% & 93.84\% & 94.89\% \\
\multicolumn{1}{l|}{beta+perturb}           & \cmark       & \xmark   & \multicolumn{1}{l|}{BETA} & 85.63\% (76.91\% - 93.43\%) & 64.92\% & 79.83\% & 86.50\% & 89.23\% & 92.63\% & 93.85\% & 94.67\% \\
\multicolumn{1}{l|}{beta+perturb+syn}       & \cmark       & \cmark   & \multicolumn{1}{l|}{BETA} & 86.71\% (76.44\% - 93.74\%) & 64.47\% & 80.16\% & \bf{88.00\%} & 89.96\% & 92.62\% & 93.77\% & 94.52\% \\ \bottomrule
\end{tabular}
\end{adjustbox}
\end{table*}
Before we analyze the FPR networks, the frozen CG framework achieved $100\%$ sensitivity on the In-house Benchmark and $97.71\%$ sensitivity on the NLST Benchmark when having 100 average candidates per scan.
We summarize the FROC curves for benchmarking the nodule detection FPR models trained with different strategies in Table~\ref{tab:froc}. Similar LUNA16 \cite{luna16}, the final CPM score is defined as the average sensitivity at 7 log-scaled false positive rates: 0.125, 0.25, 0.5, 1, 2, 4, and 8 FPs per scan. The performance obtained by training the DeepLung\footnote{https://github.com/wentaozhu/DeepLung} \cite{zhu2018deeplung} (55.15\% CPM and 75.71\% CPM) is used as a reference standard for the detection performance for the baseline detection frameworks on our benchmark datasets. Notably, our In-house benchmark has larger variances in nodule types, sizes and contexts than the LIDC dataset \cite{lidc}. The classification head with beta distribution (baseline-beta) produced similar CPM scores as the sigmoid head trained with binary cross-entropy (baseline-ce). However, we also show that the classifier would generate slightly higher CPM scores if the network is firstly trained with cross-entropy and then finetuned with the beta-distribution loss (baseline-beta-finetune).
In the experiments beta-syn (random) and beta+syn, we respectively added synthetic nodules randomly sampled from the standard normal, and the ones searched using the proposed PGD sampling. Though both types of synthetic nodules can improve the overall network generalization, the nodules searched with PGD consistently outperforms its counterpart, especially at the region of the lower number of false positives.
We show that adding the noise perturbation augmentation patches (beta+perturb and beta+perturb+syn) can also slightly improve the overall CPM scores comparing to the conventional training baseline (baseline-beta-finetune). However, they do not show better performance than only using only PGD searched nodules (beta+syn). We show such perturbation augmented networks are more robust to both uniform and adversarial noise in the next section.
The data augmentation by adding random Poisson noise \cite{huang2018some} into the training patches (beta+poisson and beta+poisson+syn) achieved slightly higher detection performance than adding the adversarial noise in both clean benchmarks. The complexities of the detection models are summarized in Table~\ref{tab:complexity}.

\begin{figure}[!htb]
     \centering
     \begin{subfigure}[b]{0.45\textwidth}
         \centering
         \includegraphics[width=\textwidth]{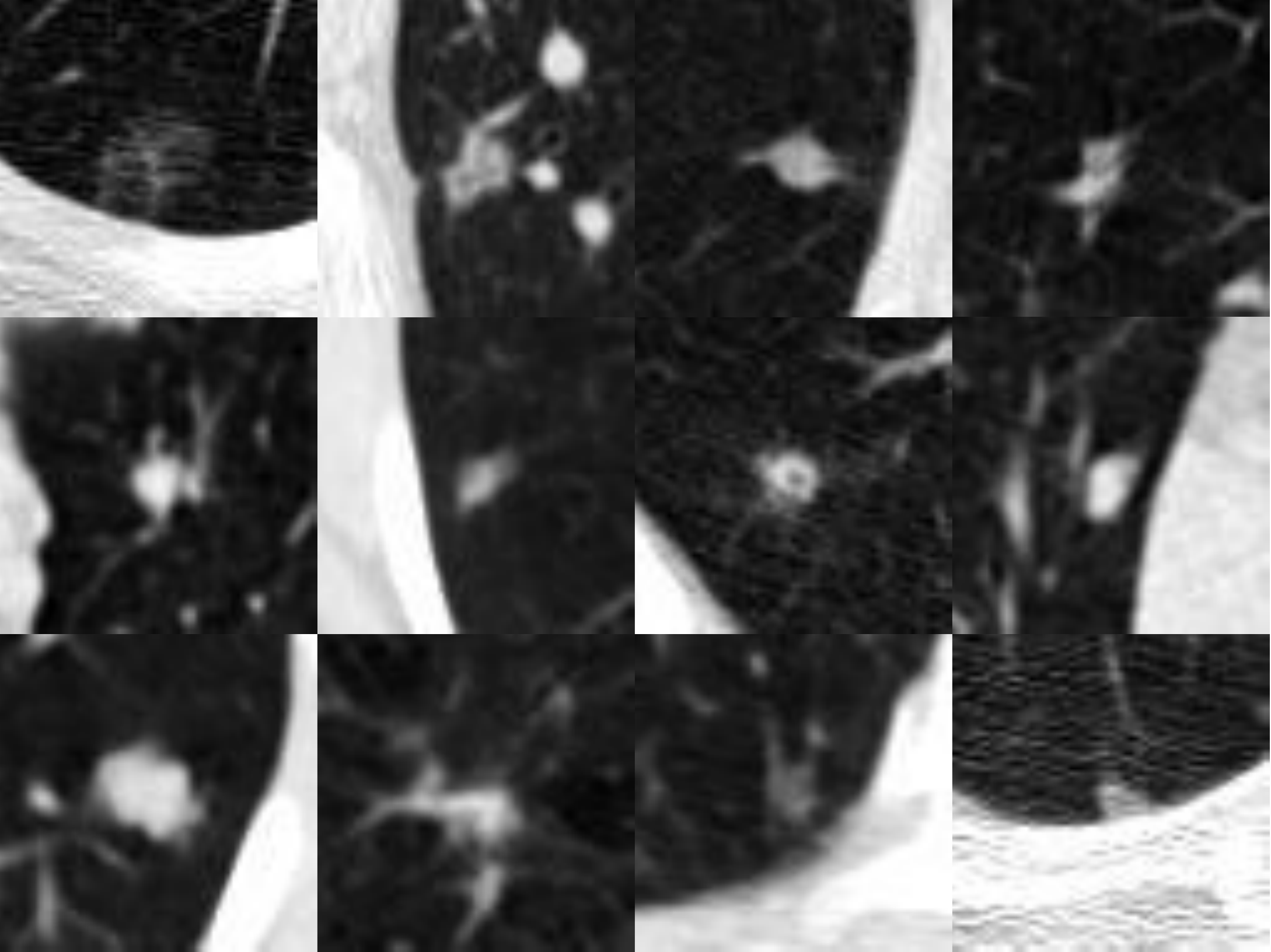}
         \caption{Real nodules}
         \label{fig:random_syn_1}
     \end{subfigure}
     \hfill
     \begin{subfigure}[b]{0.45\textwidth}
         \centering
         \includegraphics[width=\textwidth]{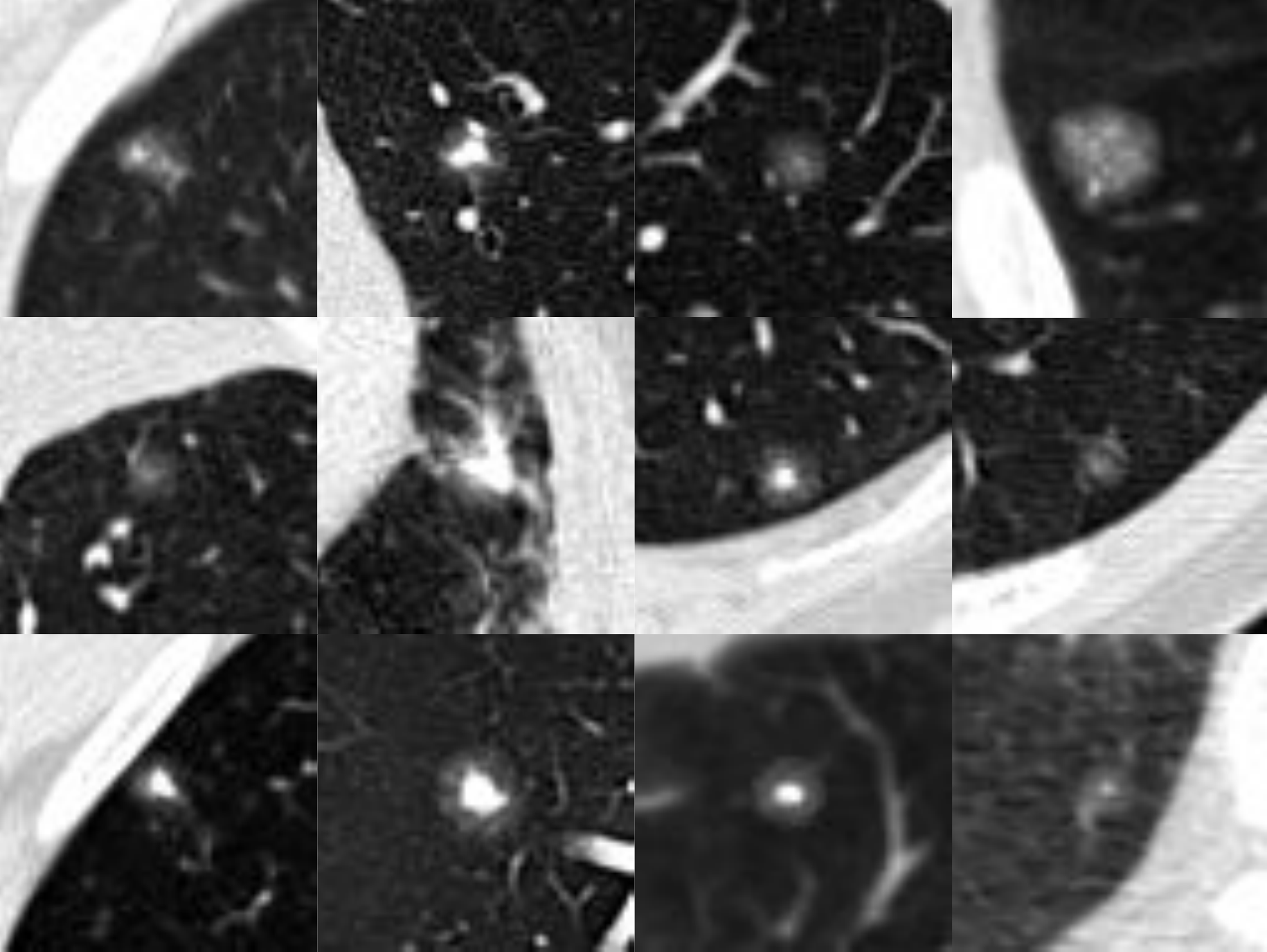}
         \caption{PGD synthetic nodules}
         \label{fig:pgd_syn_1}
     \end{subfigure}
        \caption{The mosaic view to compare the real nodule patches and the synthetic nodules in patches of size $64^3$ and $0.625^3$mm resolution. Besides being generally smaller, the PGD searched synthetic nodules tend to have round glass component with or without a solid core. Such non-solid or part-solid nodules are relatively rare in the real datasets.}
        \label{fig:real_vs_syn}
\end{figure}

\begin{table}[]
\caption{False positive reduction confidence means and standard deviations obtained on the synthetic nodule generated by using randomly sampled coding (Randomly sampled nodules) and the PGD sampled coding (Randomly sampled nodules).}
\label{tab:stress_syn_nodules}
\begin{tabular}{@{}lll@{}}
\toprule
Method                 & Randomly sampled nodules    & PGD sampled nodules       \\ \midrule
baseline-beta-finetune & $0.82 \pm 0.17$ & $0.53 \pm 0.15$ \\
beta+syn               & \boldmath{$0.92 \pm 0.09$} & \boldmath{$0.88 \pm 0.16$} \\ \bottomrule
\end{tabular}
\end{table}


\begin{figure*}[!htb]
    \centering
    \includegraphics[width=1\linewidth]{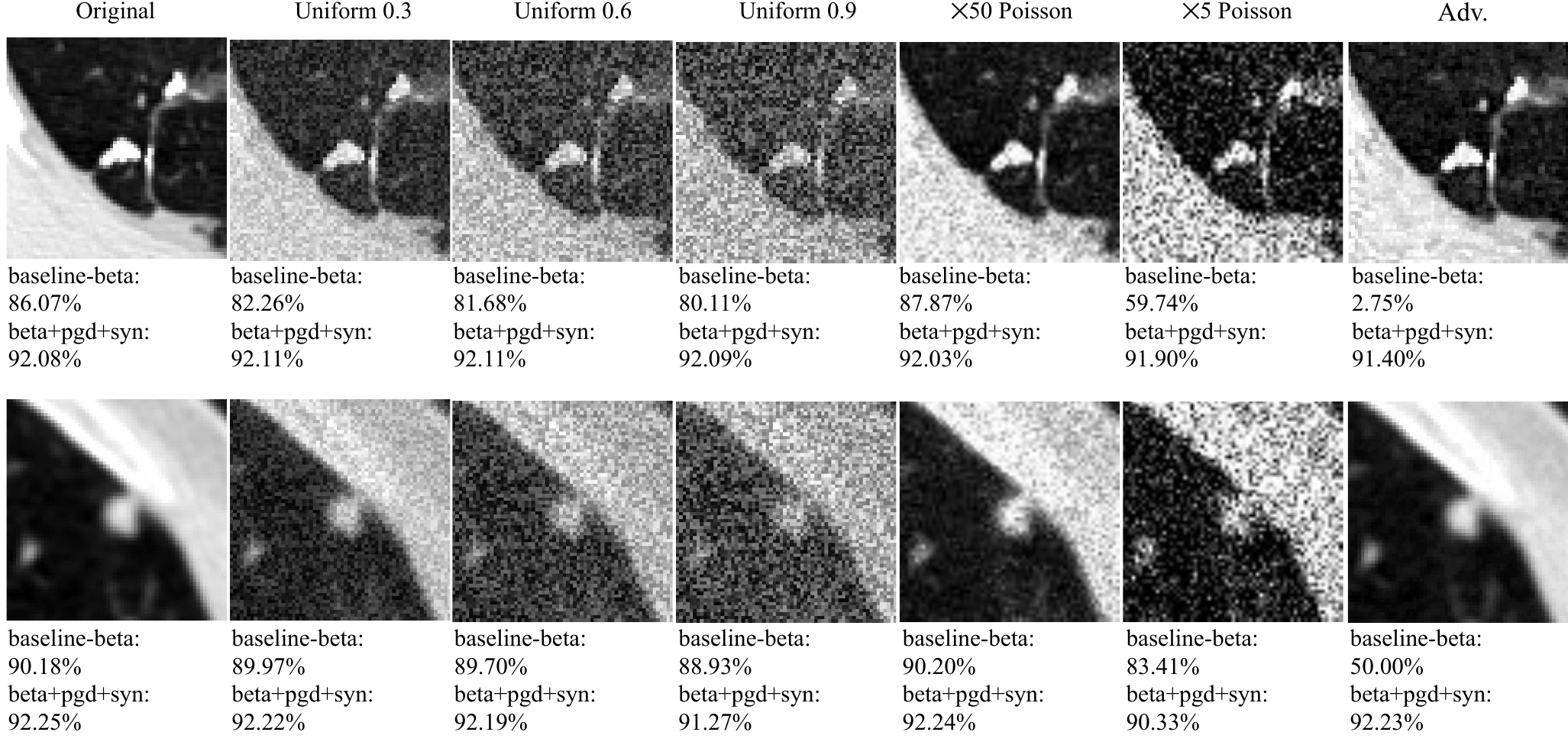}
    \caption{Examples to show different levels of uniform noise and adversarial noise on two nodules randomly drawn from the stress-test.}
    \label{fig:noised_positive_response}
\end{figure*}

\subsection{Stress test}
\subsubsection{Synthetic nodules}
The central slices of the randomly selected real nodules and the hard nodules sampled by PGD are shown in Fig.~\ref{fig:real_vs_syn}. Besides being generally smaller, the PGD searched synthetic nodules tend to have round glass component with or without a solid core. Such non-solid or part-solid nodules are relatively rare in the real datasets.
Though one can still visually distinguish a subset of the synthetic nodules from the real nodules, they can be a valuable source to stress-test the FPR network as most of such cases reside at the original decision boundaries.
We synthesized 10000 nodules with both random Gaussian sampling and PGD searching respectively. They were fed to the FPR networks trained with (beta+syn) and without (baseline-beta-finetune) synthetic nodules. We ensured that all the synthetic nodules in this test have diameters at least 6mm.
The mean and standard deviations of the network responses are shown in Table. ~\ref{tab:stress_syn_nodules}. Though the conventional network achieved $88.90\%$ CPM, it failed to recognize many nodules generated with only random sampling synthesis.
The baseline network predicts the majority of the PGD searched nodules around 0.53, which are defined as out-of-distribution (OOD) samples.
The network augmented with PGD synthetic nodules can successfully recognize most of the PGD synthetic nodules with high-confidence with mean confidence 0.87.

\subsubsection{Noise}
To stress-test the network resistance to different levels of noise, we first add uniform noise with different magnitudes to the nodule patches as depicted by Fig.~\ref{fig:noised_positive_response}.
The uniform noise can reduce the baseline network response from 0.86 to 0.81 as shown in Table.~\ref{tab:stress_positive_noise}.
We found that the network augmented with either synthetic nodules (beta+syn) or PGD noises (beta+perturb and beta+perturb+syn) can be more robust to uniform noise. 
To simulate the Poisson noise in CT, we rescaled the CT patches to [0, 50] and [0, 1] respectively and then sample from them following the Poisson process. Similarly to the uniform noise, stronger ($\times$5) Poisson noise can deactivate the baseline FPR network (from 0.86 to 0.78) while affecting less on the augmented networks augmented with PGD noise. Interestingly, the networks augmented with Poisson noise (beta+poisson and beta+poisson+syn) did not show much better robustness towards Poisson noise than the baseline network.
The networks trained without adversarial noise augmentation tend to be deactivated by the adversarial noise. At the same time, the mean responses from the adversarial noise augmented networks remained around 0.9.
We also tested the FPR network by feeding randomly generated noise patches and PGD adversarial noise patches as shown in Table.~\ref{tab:stress_negative}. The conventional FPR network would normally not be activated by random uniform noise, meaning most of the mean responses are below $0.22$. However, pure adversarial noise patches can easily activate the baseline networks (0.87 and 0.90). The networks augmented with the Poisson noise show stronger robustness than the baseline networks (0.23 and 0.34). The networks augmented by the adversarial noise augmentation (beta+perturb and beta+perturb+syn) are more robust to both types of noise patterns with below $0.1$ mean responses. In Table.~\ref{tab:stress_negative} we also show that the networks augmented with either the adversarial noise and Poisson noise are more robust to the adversarial noise added to the real negative CT patches than the baseline network.
\begin{table*}[]
\caption{Stress-test by perturbing the positive patches with different levels of uniform, Poisson noise and PGD noise perturbation.}
\label{tab:stress_positive_noise}
\begin{adjustbox}{width=\textwidth,center}
\begin{tabular}{@{}llllllll@{}}
\toprule
Method                 & 0.0 uniform noise          & 0.3 uniform noise           & 0.6 uniform noise                & 0.9 uniform noise          & $\times$50 Poisson noise   & $\times$5 Poisson noise    & Adv.       \\ \midrule
baseline-beta-finetune & $0.86 \pm 0.12$            & $0.86 \pm 0.11$             & $0.83 \pm 0.14$                  & $0.81 \pm 0.16$            & $0.86 \pm 0.11$            & $0.78 \pm 0.18$            & $0.25 \pm 0.25$ \\
beta+syn               & $0.91 \pm 0.10$            & $0.91 \pm 0.10$             & $0.89 \pm 0.11$                  & $0.87 \pm 0.16$            & $0.91 \pm 0.09$            & $0.84 \pm 0.19$            & $0.37\pm 0.39$  \\
beta+perturb           & $0.91 \pm 0.06$            & $0.91 \pm 0.05$             & \boldmath{$0.91 \pm 0.05$}       & \boldmath{$0.89 \pm 0.12$} & $0.91 \pm 0.05$            & \boldmath{$0.91 \pm 0.07$} & \boldmath{$0.91 \pm 0.07$}      \\
beta+perturb+syn       & \boldmath{$0.92 \pm 0.04$} & \boldmath{$0.91 \pm 0.04$}  & $0.90 \pm 0.07$                  & $0.87 \pm 0.13$            & \boldmath{$0.92 \pm 0.04$} & $0.90 \pm 0.07$            & $0.90 \pm 0.08$     \\
beta+poisson           & $0.86 \pm 0.07$            & $0.86 \pm 0.07$             & $0.84 \pm 0.09$                  & $0.83 \pm 0.11$            & $0.86 \pm 0.07$            & $0.80 \pm 0.14$            & $0.78 \pm 0.18$     \\
beta+poisson+syn       & $0.90 \pm 0.07$            & $0.89 \pm 0.09$             & $0.88 \pm 0.11$                  & $0.86 \pm 0.14$            & $0.89 \pm 0.07$            & $0.79 \pm 0.21$            & $0.82 \pm 0.17$     \\ \bottomrule
\end{tabular}
\end{adjustbox}
\end{table*}

\begin{table}[]
\caption{Stress test by feeding noise patches and the negative patches to the network.}
\label{tab:stress_negative}
\begin{adjustbox}{width=\linewidth,center}
\begin{tabular}{@{}lllll@{}}
\toprule
Method                 & Uniform noise & PGD noise & Negative patches & PGD negative patches \\ \midrule
baseline-beta-finetune & $0.11 \pm 0.05$        & $0.87 \pm 0.17$    & $0.14 \pm 0.09$      & $0.82 \pm 0.21$  \\
beta+syn               & $0.22 \pm 0.16$        & $0.90 \pm 0.10$    & \boldmath{$0.10 \pm 0.11$}      & $0.63 \pm 0.32$ \\
beta+perturb           & \boldmath{$0.04 \pm 0.00$}        & \boldmath{$0.08 \pm 0.13$}    & $0.19 \pm 0.14$      & $0.26 \pm 0.20$ \\
beta+perturb+syn       & $0.05 \pm 0.01$        & $0.09 \pm 0.09$    & $0.21 \pm 0.15$      & $0.32 \pm 0.23$  \\
beta+poisson           & $0.13 \pm 0.06$        & $0.23 \pm 0.21$    & $0.13 \pm 0.12$      & $0.18 \pm 0.18$  \\
beta+poisson+syn       & $0.08 \pm 0.02$        & $0.34 \pm 0.25$    & $0.11 \pm 0.10$      & \boldmath{$0.17 \pm 0.17$} \\ \bottomrule
\end{tabular}
\end{adjustbox}
\end{table}

\section{Discussions and Conclusions}
In this paper, we propose adversarial augmentation methods to improve the robustness of the nodule detection framework against two major sources of the out-of-distribution samples (1) the nodules with under-represented properties in the training dataset (2) the images with unexpected noise or contrast.
We first use the beta-distribution to replace the sigmoid output of the false positive reduction network to estimate the observation uncertainty explicitly at the output layer.
Then we add both adversarial synthetic nodules and adversarial perturbation noise to the training set that is searched using the project gradient descent (PGD).
Some of the existing works as listed in S.II.B have attempted to use synthetic images to improve the model performance since generative models can provide samples under-represented in the training set which is not easily simulated by the conventional data augmentation techniques. However, as shown in Table II, the performance gain using randomly sampled synthetic images is limited since the majority of the randomly sampled cases drawn from a prior distribution are well represented by the training data distribution. Therefore, we propose to search for the latent code associated with the hard-samples from the synthesizer using PGD to maximize the training loss of the supervised model. This applies when (1) the data synthesizer and the supervised learning model are both differentiable (2) the data synthesizer is trained to well-represent the manifold of interests. 
We show that the proposed techniques can improve the generalization of the nodule detector by learning pathologically relevant patterns, we tested it on two benchmark datasets with groundtruth annotated by experienced radiologists.
We also use the synthetic nodules and the generated perturbations to stress test the trained models and show the augmented networks can be more robust to both hard nodules as well as different types of noise distortions.
By using the beta distribution based uncertainty estimation, we also showed that uncertainty estimation alone might not be sufficient to make the network robust to the out-of-distribution inputs, especially when the inputs are adversarially generated.
The proposed framework can be indeed applied to similar classification, segmentation or detection models as long as the model and the data synthesizer are fully differentiable.

Some studies observed that adversarial samples can cause decreased testing performance on clean testing data~\cite{stutz2019disentangling,2018arXiv180512152T,raghunathan2019adversarial,xie2020intriguing}.
However, bearing in mind that dropped accuracy can be expected as a cost for robustness, in our experiments shown in Table.~\ref{tab:froc}, we did not notice obvious testing performance decreases on the clean data though the best performance is indeed observed on the models with only adversarial hard-case sampling without adversarial noise. This observation is also consistent between the two benchmark datasets. Our assumptions are two-fold: (1) As hypothesized in \cite{xie2020intriguing}, the generalization performance drop might be because the clean data and the adversarial data are drawn from different manifolds. Even though detecting rare nodules automatically has been a challenging problem for decades, the underlying data manifolds of the nodules can be much simpler than the ones of natural image applications. The neural networks can, therefore, be over-parameterized to fit both the clean and the adversarial manifold without changing the normalization layers. (2) Other than adversarial noise, the adversarial noise enhanced networks can be also more robust to other noise or artifact sources that can appear in clean CT data. However, these assumptions might need to be further validated by future work.

As one of the early attempts to enhance the robustness of the medical image analysis CNNs, this study has a few limitations that would be targeted in future works.
We use a relatively simple nodule synthesizer network and the standard PGD to sample the lung nodules from the latent space. This synthesizer was not capable of synthesizing all types of different nodules, such as nodules with spiculation.
It was also not constrained to maintain the size of a synthetic nodule, therefore we had to filter out the synthetic nodules that are smaller than the relevant threshold.
We only investigated the network robustness towards three types of image noise. The improved robustness towards other types of image artefacts, such as metal artefacts and motion distortion, etc., remains unknown.
As a proof of concept study, the proposed techniques were only applied to the false positive reduction (FPR) of the lung nodule detection pipeline for brevity. However, the same perturbations can also affect candidate generation networks. 
We also found that in practice it is hard to generate adversarial noise by attacking the noise augmented networks without showing visually detectable artefacts.
However, it is possible to attack augmented networks with the same techniques.
Though we only evaluated the proposed techniques in the context of nodule detection, we believe such techniques can also be helpful for the other deep CNN based medical imaging applications with minor technical adjustments.

\noindent\textbf{Disclaimer}: The concepts and information presented in this paper are based on research results that are not commercially available 

\bibliographystyle{IEEEtran}
\bibliography{refs.bib}{}

\end{document}